\documentclass[12pt]{article}

\usepackage{newtxtext,newtxmath}

\usepackage{graphicx}

\usepackage[letterpaper,margin=1in]{geometry}
\usepackage{longtable}
\usepackage{tabularx}
\usepackage{amsmath}
\usepackage{mathtools}
\usepackage{float}
\usepackage{booktabs}
\usepackage[most]{tcolorbox}
\makeatletter
\let\amsthm@old@openbox\openbox
\let\openbox\relax
\usepackage{amsthm}
\let\openbox\amsthm@old@openbox
\makeatother
\renewenvironment{abstract}
	{\quotation}
	{\endquotation}
\date{}

\makeatletter
\renewcommand{\fnum@figure}{\textbf{Figure \thefigure}}
\renewcommand{\fnum@table}{\textbf{Table \thetable}}
\makeatother

\usepackage{scicite}

\usepackage{url}

\def\scititle{
Quantum Statistics Forbids Particle Exchange Statistics beyond Bosons and Fermions in 3D
}
\title{\bfseries \boldmath \scititle}

\author{
    Chi-Chun Zhou$^{1\dagger*}$, 
    Shuai A. Chen$^{2\dagger\#}$, 
    Yu-Zhu Chen$^{3}$, 
    Yao Shen$^{4}$, \\
    Fu-Lin Zhang$^{5}$, 
    Wu-Sheng Dai$^{5**}$\and
	\small$^{1}$School of Engineering, Dali University, Yunnan 671003, China.\and
	\small$^{2}$ Max Planck Institute for the Physics of Complex Systems, Nöthnitzer Straße 38, Dresden 01187, Germany.\and
    \small$^{3}$School of Physical Science and Technology, Tiangong University, Tianjin 300387, China. \and
    \small$^{4}$School of Criminal Investigation, People's Public Security University of China, Beijing 100038, China.\and
    \small$^{5}$Department of Physics, Tianjin University, Tianjin 300350, China. \\
    \small$^{\#}$  Email: chsh@pks.mpg.de   \\
    \small$^{\ast}$  Email: zhouchichun@dali.edu.cn   \\
	\small$^{\ast\ast}$  Email: daiwusheng@tju.edu.cn  \\
	\small$^\dagger$These authors contributed equally to this work. \\
}

\begin{document} 

\maketitle

\begin{abstract}\bfseries \boldmath
\noindent
Quantum matter in three spatial dimensions is observed to consist exclusively of bosons and fermions. Whether this empirical fact follows from basic consistency requirements of quantum theory itself or must be imposed as an additional principle has for 80 years remained a fundamental conceptual gap. Here we close this gap by establishing a no-go theorem that excludes any particle exchange statistics beyond bosons and fermions in three dimensions. We identify the consistency conditions linking the many-body Hilbert-space structure of quantum mechanics with the statistical microstate counting of indistinguishable particles. As a corollary, we demonstrate that higher-dimensional representations of the symmetric group cannot give rise to genuinely distinct particle exchange statistics in any spatial dimension. 
\end{abstract}

\noindent
\section*{Introduction}
Quantum matter in three spatial dimensions is empirically observed to consist only of bosons and fermions \cite{shi2018experimental,kaplan2020pauli,english2010spectroscopic,ramberg1990experimental,bellini2010new}. In two dimensions, the topology of configuration space allows for anyonic exchange statistics \cite{wilczek1982quantum,leinaas1977theory,Laidlaw1971,wu1984general,bartolomei2020fractional,nakamura2020direct}.
A long-standing foundational question remained: Does the exclusion of particle exchange statistics beyond bosons and fermions in three spatial dimensions follow from the theory of quantum mechanics, or must it be imposed as an additional postulate? This uncertainty has persisted as a fundamental conceptual gap in our understanding of quantum statistics.

The search for a “third kind” of particle exchange statistics has doggedly continued for over eight decades, giving rise to two conceptually distinct lines of inquiry. One line of work seeks to derive restrictions on exchange statistics directly from quantum mechanics \cite{doplicher1969fields,doplicher1971local,doplicher1974local,messiah1964symmetrization,girardeau1965permutation,flicker1967symmetrization,bros1966theoretical,Laidlaw1971,leinaas1977theory,BerryRobbins1997,Balachandran1993,balachandran1991classical}. These studies typically invoke additional structural assumptions—such as relativistic locality, microcausality, or explicit symmetrization constraint. 
While  compelling within their respective settings, they do not directly apply to a broad class of physically relevant systems, such as lattice models and strongly correlated platforms. 

A complementary line of work adopts a constructive perspective by proposing explicit “intermediate” statistics interpolating between bosons and fermions. Examples include finite-occupancy rules \cite{gentile1940itosservazioni,dai2004gentile}, exclusion-based counting schemes \cite{haldane1991fractional,wu1994statistical}, higher-dimensional representations of the symmetric group \cite{green1953generalized,greenberg1990example,okayama1952generalization,tichy2017extending}, modified exchange algebras \cite{greenberg1991particles,greenberg1993quons,cattani1984general,ohnuki1982quantum}, and formulations based on symmetric functions \cite{zhou2022unified,schmidt2002partition,chaturvedi1997interpolations}. Those constructions, however, are known to fail to yield a new form of particle exchange statistics, consistent with the empirical absence of a “third statistics.” 
Thus, a general  criterion is needed to assess the  proposed statistics scheme. Recent claims of new paraparticles beyond bosons and fermions \cite{wang2025particle} further underscore the need for such a systematic consistency analysis.

In this work, we establish a no-go theorem that definitively rules out any exchange statistics beyond bosonic and fermionic types in three spatial dimensions.\\ 
\textbf{No-go theorem.}
For an ideal system of identical particles in three spatial dimensions, where particle exchanges are governed by the symmetric group, any proposed generalization beyond bosonic and fermionic statistics necessarily fails as a `third kind' of quantum statistics. Specifically, such a construction must either (i)  break the microstate counting of indistinguishability or (ii) lack a well-defined quantum many-body Hilbert space. \\
Most importantly, our no-go theorem determines a consistent condition between quantum mechanics and statistical mechanics, which has not been fully discussed before. 
Regarding quantum statistics, the two aspects, quantum mechanics and statistical mechanics,  are inseparable: physically meaningful quantum statistics must both admit a well-defined many-body Hilbert space realization \emph{and} obey the microstate counting for indistinguishable particles. When these two requirements are imposed together in three spatial dimensions, only bosonic and fermionic statistics remain viable.

\section*{Results}
When we try to construct a “third type” of particle exchange statistics, we can consider the  canonical partition function, which bridges between quantum mechanics and  statistical mechanics \cite{reichl2016modern,pathria2017statistical}: 
\begin{equation}
Z(\beta,N)=\mathrm{Tr}_{\mathcal H_N}\!\left(e^{-\beta H}\right)
=\sum_E \Omega(E,N)\,e^{-\beta E},
\label{eq:Z}
\end{equation}
where the Hamiltonian $H$ acts on the $N$-particle Hilbert space $\mathcal H_N$, and $\beta=(k_B T)^{-1}$ with $k_B$ the Boltzmann constant and $T$ the temperature. $\Omega(E,N)$ denotes the number of microstate at fixed energy $E$ and particle number $N$. 
The left equality in Eq.~\eqref{eq:Z} speaks the language of quantum mechanics (trace over a many-body Hilbert space with Hamiltonian),
while the right equality speaks the language of statistical mechanics (count and sum over microstates).
Genuine particle exchange statistics must make these two languages describe the \emph{same} $Z(\beta,N)$.


\textbf{Perspective of quantum mechanics}. 
In three dimensions, where all particle-exchange loops are contractible, the indistinguishability requires that the many-body wave functions transform according to representations of the symmetric group $S_N$. For instance, bosons and fermions correspond to one-dimensional representations of the symmetric group $S_N$. 
Accordingly, the many-particle Hilbert space is constructed by taking tensor products of the single-particle Hilbert space and organizing the resulting space into representations of $S_N$. Simultaneously, 
the many-particle Hilbert space should be invariant regardless of the basis of the single-particle Hilbert space when we take a tensor product. It reflects the invariance of the inner product of Hilbert space under the single-particle basis transformation, which forms a $U(m)$ group. Such a structure of the many-body Hilbert space can be characterized using the Schur-Weyl duality \cite{weyl1946classical}, 
\begin{equation}
 (\mathbb C^m)^{\otimes N}= \bigoplus_{I=1}^{P(N)} \left(V_{U(m)}^I \otimes V_{S_N}^I\right),
\end{equation}
where $V_{U(m)}^I$ and $V_{S_N}^I$ denote irreducible representations of $U(m)$ and $S_N$, respectively. These irreducible components are uniquely labeled by integer partitions $(\lambda)_I$ of $N$. $P(N)$ counts the number of the integer partitions of $N$. Therefore, the Schur-Weyl duality determines that a physical many-body Hilbert space $\mathcal H_N$ carries the irreducible representations of the $U(m)$ group.
 
We are directed to an expression of the  canonical partition function  $Z(\beta,N)$ in Eq.~(\ref{eq:Z}) \cite{zhou2022unified,chaturvedi1997interpolations,chaturvedi1997microscopic} with details can be found in the Supplementary Information,
\begin{equation}
Z(\beta ,N)\;=\sum_{I=1}^{P(N)}C^{I}\,s_{I}\!\left(
x_{1},x_{2},\cdots,x_m \right) \,,\label{eq:Zs}
\end{equation}
where $s_{I}(x_1,x_2,\cdots,x_m)$ is the Schur polynomial associated with $x_{i}=e^{-\beta \epsilon_i}$. By noting \(C^{I}\) counts the multiplicity with which the irrep \(V_{U(m)}^{I}\) appears in the Hilbert space $\mathcal H_N$.
which immediately yields the first constraint:
\begin{equation}
\text{The multiplicities } C^{I} \text{ must be non-negative integers:}
\qquad
C^{I} \in \mathbb{N}_0 \quad \forall I .
\label{constrain1}
\end{equation}
If Eq.~\eqref{constrain1} is violated, the resulting space cannot be interpreted as a Hilbert space, and the system therefore falls outside the framework of quantum mechanics.
Thus, one attempts to design intermediate exchange statistics simply by selecting an allowed symmetry sector,
by imposing $C^{I}>0$ for a selected symmetry sector while $C_I = 0$ for others, and thereby defining a Hilbert-space structure.

\textbf{Perspective from statistical mechanics}.
To formulate the statistical constraint, we reinterpret the partition function in Eq.~\eqref{eq:Z} from a combinatorial perspective. 
Instead of the ensemble angle in Eq.~\eqref{eq:Z} which we are more familiar with, we can reformulate the microstate counting of indistinguishable particles through the partitions of an integer $N$. We can group all the microstates according to the partition of $N$: $(\lambda)_J \equiv (\lambda_1, \lambda_2, \cdots,\lambda_m) $ for $J=1,\cdots ,P(N)$, and each partition corresponds to a set of occupation configurations. 
Then, a canonical partition function is expressed as \cite{zhou2022unified,chaturvedi1997microscopic},
\begin{equation}
Z(\beta,N)
=\sum_{J=1}^{P(N)} \Omega^{J}\,
m_{J}\!\left(x_{1},x_{2},\ldots\right),
\label{eq:Zm}
\end{equation}
where $m_J\left(x_{1},x_{2},\ldots\right)$ is the monomial symmetric function \cite{macdonald1998symmetric}
associated with the occupation-number pattern $(\lambda)_J$ (the $J$th integer partition of $N$).
For indistinguishable particles, the requirement that relabeling particles does not generate a new microstate manifests as the second constraint:  
\begin{equation} \label{constrain2}
    \text{The coefficents $\Omega^J$ take binary values:} \quad \Omega^J\in \{0,1\} \quad \forall J.
\end{equation}
The constraint reflects that $\Omega^{J}$ functions as an indicator, taking the value $1$ only when the configuration $(\lambda)_K$ satisfies the statistical conditions on state occupations, such as the Pauli exclusion principle for fermions, and $0$ otherwise. If the condition in Eq.~\eqref{constrain2} is violated, for example, for the Maxwell–Boltzmann statistics, the $\Omega^J>1$, indicating microstate counting of distinguishable particles. 
Furthermore, from Eq.~\eqref{eq:Zm}, one possible generalization is to specify some $\Omega^J$s equal to $1$.

\subsection*{No-go theorem}
Up to now, we have derived two formulas from the perspectives of quantum mechanics and statistical mechanics regarding the partition function. 
In what follows, we examine the conditions under which these two constraints are mutually consistent.
Equivalently, the two expressions in Eqs.~\eqref{eq:Zm} and \eqref{eq:Zs} are required to represent the same partition function.

These formulas, Eqs.~\eqref{eq:Zm} and \eqref{eq:Zs}, involve two classes of symmetric polynomials.
Mathematically, they are related by a linear transformation whose coefficients are given by the integer-valued Kostka numbers $k_I^J$ \cite{littlewood1977theory,macdonald1998symmetric} (details can be found in the Supplementary Information):
\begin{equation}\label{eq:kij}
  \Omega^{J} = \sum_{I=1}^{P(N)} k^{J}_{I}\, C^{I}\,.
\end{equation}
Eq.~\eqref{eq:kij} encodes the consistency condition between the many-body Hilbert-space construction and the microstate counting of indistinguishable particles. As shown in the Methods section, the resulting Kostka-number constraints admit only two consistent solutions, corresponding to Bose and Fermi statistics. This establishes our no-go theorem.

The above no-go result in fact extends to any spatial dimension.\\
\noindent\textbf{Corollary.}—
For indistinguishable particles in any spatial dimension, any attempt to realize quantum statistics using higher-dimensional representations of the symmetric group $S_N$ fails to produce a genuinely new kind of quantum statistics.

\begin{table}[t]
\centering
\caption{Representative no-go theorems for quantum statistics in three dimensions.
``QM'' denotes the existence of a well-defined quantum-mechanical Hilbert space.
``SM'' denotes consistent microstate counting for indistinguishable particles from the perspective of statistical mechanics.
``QS'' requires that both conditions be satisfied simultaneously.
We emphasize that our assessment concerns whether a given scheme constitutes fundamental quantum statistics. These schemes can arise as effective descriptions of low-energy systems. 
Here, $\times$/$\checkmark$ means that the case is parameter-dependent.}
\label{tab:qsit-logic}
\vspace{6pt}
\begin{tabular}{lccc}
\toprule
 & \textbf{QM} & \textbf{SM} & \textbf{QS} \\
\midrule
Boson                    & $\checkmark$ & $\checkmark$ & $\checkmark$ \\
Fermion                  & $\checkmark$ & $\checkmark$ & $\checkmark$ \\
\midrule
Gentile (1940) \cite{gentile1940itosservazioni}            & $\times$     & $\checkmark$ & $\times$ \\
Green's parastatistics (1953) \cite{green1953generalized}     & $\checkmark$ & $\times$     & $\times$ \\
Biedenharn--Macfarlane (1989) \cite{biedenharn1989quantum}   & $\times$     & $\times$/$\checkmark$ & $\times$ \\
Haldane--Wu (1991) \cite{haldane1991fractional}             & $\times$     & $\times$     & $\times$ \\
Greenberg's quon (1993) \cite{greenberg1993quons}          & $\checkmark$ & $\times$     & $\times$ \\
Jack-polynomial schemes (1997) \cite{chaturvedi1997interpolations}  & $\times$/$\checkmark$ & $\times$ & $\times$ \\
Immanons  (2017) \cite{tichy2017extending}              & $\checkmark$ & $\times$     & $\times$ \\
Parastatistics (2025) \cite{wang2025particle}   & $\times$     & $\times$     & $\times$ \\
\bottomrule
\end{tabular}
\end{table}

\section*{Discussion}
The absence of particle statistics beyond bosons and fermions in three spatial dimensions is firmly established experimentally.
We demonstrate that such statistical constraint does not stand as an independent postulate, and it can be derived regarding the particle indistinguishability when particle exchanges are governed by the symmetric group $S_N$. Table~\ref{tab:qsit-logic}, we apply our no-go theorem to some representative generalizations for quantum statistics.

In general, generalizations for new quantum statistics follow two strategies: realizing high-dimensional representations of the symmetric group, or imposing alternative occupation rules to define generalized statistics. 
As summarized in Table~\ref{tab:qsit-logic}, constructions based on higher-dimensional representations of the symmetric group $S_N$ include parastatistics and related approaches \cite{green1953generalized,tichy2017extending,greenberg1990example,cattani1984general,ohnuki1982quantum,okayama1952generalization,chaturvedi1996canonical}. 
These schemes 
inevitably break the microstate counting of indistinguishable particles, which fail as a proper statistical mechanical system. 
In Green’s parastatistics \cite{green1953generalized}, for instance, parabosons and parafermions admit consistent many-body Hilbert-space realizations, but their statistical mechanics acquires an internal multiplicity factor $\Omega^J>1$ that reflects extra internal degrees of freedom. An important advantage of our no-go theorem is that it does not rely on measurement. Instead, it bypasses ambiguities associated with whether such internal degrees of freedom are observable, and directly diagnoses the inconsistency at the level of many-body state counting. In this sense, our no-go theorem is more applicable regarding the generalizations like Green's parastatistics.
Meanwhile, schemes that impose restrictions on admissible occupation patterns—such as Gentile statistics \cite{gentile1940itosservazioni,dai2004gentile} can stastify  the microstate counting for indistinguishable particles, but do not admit a well-defined many-body Hilbert space. These schemes may provide effective descriptions of low-energy excitations with emergent statistical behavior. Nevertheless, our results indicate that they do not represent fundamental particle exchange statistics.

\section*{Outlook}
An important and promising direction is to move beyond the symmetric-group paradigm by considering systems in which particle exchange is governed by the braid group. Two-dimensional anyonic systems, for example, lie outside the scope of our no-go theorem, and many-body states depend on additional topological data, such as fusion channels and braiding histories, beyond simple occupation patterns. Extending the framework of our no-go theorem to such braid-group–based settings could clarify which types of topological constraints permit genuinely new forms of exchange statistics beyond Bose and Fermi. More broadly, this approach may provide insight into the rigidity or possible generalizations of quantum statistics in systems with nontrivial topology.

\section*{Methods}

The no-go theorem follows from the compatibility between the quantum-mechanical and statistical-mechanical representations of the canonical partition function, which leads to the Kostka-number relation in Eq.~\eqref{eq:kij}. Here we summarize the essential mathematical ingredients and logic; a complete proof is provided in Supplementary Information.

The Schur polynomials $s_I$ and monomial symmetric functions $m_J$ form two bases of the space of symmetric polynomials of degree $N$ and are related by the Kostka transformation \cite{littlewood1977theory,macdonald1998symmetric},
\begin{equation}
s_I=\sum_{J=1}^{P(N)} k_I^J\, m_J ,
\end{equation}
where $k_I^J$ are the Kostka numbers \cite{littlewood1977theory,macdonald1998symmetric}. These nonnegative integers possess three key properties relevant here:  
(i) the Kostka matrix is unitriangular under the dominance ordering of integer partitions;  
(ii) all diagonal entries equal unity, $k_I^I=1$;  
(iii) all nonzero off-diagonal entries satisfy $k_I^J\ge 1$.

Equating the Schur- and monomial-basis expansions of the partition function yields the relattion
$\Omega^J=\sum_I k_I^J\,C^I $,
which must be satisfied simultaneously with the quantum-mechanical constraint $C^I\in\mathbb N_0$ and the statistical-mechanical constraint $\Omega^J\in\{0,1\}$. Because of the unitriangular structure and positivity of the Kostka matrix, these constraints can be satisfied only if exactly one coefficient $C^I$ is nonzero and equal to unity, and if all nonzero entries in the corresponding Kostka column equal one.

From the combinatorial definition of Kostka numbers, only two columns satisfy this condition: the fully symmetric and fully antisymmetric partitions. These yield Bose–Einstein and Fermi–Dirac statistics, respectively. All other choices violate either Hilbert-space integrality or the uniqueness of indistinguishable-particle microstate counting.

\section*{Acknowledgments}
\noindent This work was supported by the National Natural Science Foundation of China (62106033, 11575125, and 11675119). S.C. is grateful to the Max Planck Society for financial support. We are grateful to Zheng-Yu Weng, Benoît Douçot, Roderich Moessner, Quanhui Liu, Zhi Li, Ming Gong, Hong-Hao Tu, S. Chaturvedi, Hongwei Jia, Jing Hu, Xiaoping Ouyang, and Jixuan Hou for stimulating and helpful discussion.

\section*{Additional information}
\noindent Competing interests: The authors declare that they have no known competing financial interests or personal relationships that could have appeared to influence the work reported in this paper.

\bibliography{science_template} 
\bibliographystyle{sciencemag}

\newpage
\appendix
\renewcommand{\thesection}{SI\arabic{section}.}
\setcounter{section}{0}
\renewcommand{\thesubsection}{\Alph{subsection}}
\setcounter{subsection}{0}

\begin{center}
    \textbf{Supplementary Information for ``Quantum Statistics Forbids Particle Exchange Statistics beyond Bosons and Fermions in 3D"}
\end{center}

\tableofcontents

\section{The Motivation}

\subsection{A Long-Standing Foundational Question}
All dedicated experimental searches for violations of Bose or Fermi statistics in three-dimensional quantum systems have so far yielded null results, and are therefore consistent with bosonic or fermionic statistics \cite{shi2018experimental,kaplan2020pauli,english2010spectroscopic,ramberg1990experimental,bellini2010new}. This empirical fact finds its theoretical foundation in the spin–statistics theorem, which establishes a strict connection between particle spin and Bose/Fermi statistics under a set of well-defined structural assumptions \cite{fierz1939relativistische,1940Pauli,schwinger1958spin,luders1958connection,streater2000pct}.

Crucially, the conclusion of the spin–statistics theorem is conditional upon these assumptions, including relativistic causality and locality. Violations of the spin–statistics connection are therefore not logically excluded a priori, but can only arise at the cost of abandoning at least one of these principles \cite{greenberg1998spin}.

In two spatial dimensions, the topology of the configuration space of indistinguishable particles is fundamentally different, with particle exchanges classified by the braid group rather than the permutation group. This topological distinction allows for anyonic exchange statistics, which have been established both theoretically and experimentally \cite{wilczek1982quantum,leinaas1977theory,Laidlaw1971,wu1984general,bartolomei2020fractional,nakamura2020direct}.

The coexistence of a spin–particle connection that is contingent upon specific structural assumptions in three dimensions and the existence of anyons in two dimensions thus raises a long-standing foundational question: \textbf{whether particle exchange beyond the Bose–Fermi dichotomy can exist in three-dimensional quantum systems under the standard principles of quantum mechanics, and if not, which assumptions fundamentally forbid them}.

To address this question, research since the 1940s has developed along two broad and conceptually distinct lines.

The first line seeks to derive, or to provide alternative justifications for, the restriction of physical states to be either symmetric (bosonic) or antisymmetric (fermionic), with the aim of establishing no-go theorems that exclude any third type of particle exchange statistics. The second line explores various formally distinct models of particle exchange statistics, proposed through generalizations of statistical mechanics, quantum theory, or underlying algebraic structures.

\subsection{Existing No-Go Theorems and Their Scope}

Beyond Pauli’s original spin–statistics theorem \cite{1940Pauli}, further no-go results have been obtained by Doplicher, Haag, and Roberts \cite{doplicher1969fields,doplicher1971local,doplicher1974local}. These theorems rigorously constrain the possible exchange statistics of particles, but they remain firmly rooted in a relativistic framework.

The analogous restriction is commonly formulated as the symmetrization postulate  \cite{messiah2014quantum}, which states that for a system of $N$ identical particles, the admissible quantum states must be either totally symmetric or totally antisymmetric under the exchange of any pair of particles. Consequently, identical particles are classified as either bosons or fermions according to the symmetry of their wave functions.

However, the symmetrization postulate is widely regarded as an additional assumption rather than a consequence of the standard axioms of quantum mechanics \cite{messiah2014quantum}. For example, Messiah and Greenberg explicitly argued that the symmetrization postulate cannot be derived solely from the principles of quantum mechanics together with the indistinguishability of identical particles \cite{messiah1964symmetrization}. Motivated by this observation, a substantial body of work has sought to derive or justify the symmetrization postulate by invoking additional principles or structural assumptions \cite{girardeau1965permutation,flicker1967symmetrization,bros1966theoretical,Laidlaw1971,leinaas1977theory,BerryRobbins1997,Balachandran1993,balachandran1991classical}.

A common feature of these approaches is that they introduce, either explicitly or implicitly, further restrictions that exclude higher-dimensional representations of the permutation groups, or equivalently, forbid additional internal fibers beyond the standard quantum-mechanical degrees of freedom. While some works make these assumptions explicit, others incorporate them in a more implicit or geometric manner \cite{kaplan2013pauli,kuckert2005spin,harrison2004quantum}. For example, the analyses of Laidlaw–DeWitt and Leinaas–Myrheim already presuppose that only scalar (one-dimensional) representations of the relevant configuration-space fundamental group are physically admissible \cite{kaplan2013pauli,leinaas1977theory}.

A comparative overview of the assumptions underlying these existing no-go arguments is summarized in Table~\ref{tab:existing_nogo}.

\begin{longtable}{p{3.5cm} p{4.5cm} p{5.5cm}}
    \caption{Representative works related to the symmetrization postulate: main conclusion and key assumptions. }
    \label{tab:existing_nogo}
    \\
    \toprule
    & \textbf{Main conclusion}
    & \textbf{Key assumptions} \\
    \midrule
    \endfirsthead

    \toprule
    &
    \textbf{Main conclusion}
    & \textbf{Key assumptions} \\
    \midrule
    \endhead

    \midrule \multicolumn{3}{r}{\textit{Continued on next page}}\\
    \midrule
    \endfoot

    \bottomrule
    \endlastfoot

    Pauli's spin--statistics theorem (1940) \cite{1940Pauli, WeinbergQTF1}
    & Integer--spin fields must be bosonic; half-integer--spin fields must be fermionic
    & Relativistic quantum field theory; locality/microcausality; spectrum condition; existence and uniqueness of the vacuum
    \\
    \midrule

    Doplicher--Haag--Roberts (DHR) superselection theory \cite{doplicher1971local,doplicher1974local}
    & In 3D relativistic QFT, superselection sectors carry Bose or Fermi statistics (or parastatistics reducible to Bose/Fermi)
    & Algebraic QFT axioms (Haag--Kastler framework); relativistic covariance; locality; selection criteria for sectors; categorical analysis of statistics
    \\
    \midrule

    Laidlaw \& DeWitt (1971) \cite{Laidlaw1971}
    & In 3D, only Bose and Fermi statistics arise when restricting to one-dimensional unitary representations of the symmetry group
    & One-dimensional unitary representations of the fundamental group (symmetric group in 3D)
    \\
    \midrule

    Leinaas \& Myrheim (1977) \cite{LeinaasMyrheim1977}
    & Anyons are possible in 2D, while in 3D only Bose/Fermi statistics arise under their hypotheses
    & Scalar single-valued wavefunctions on the configuration space; no additional internal degrees of freedom; no non-Abelian bundles or non-scalar exchange representations
    \\
    \midrule

    Berry \& Robbins (1997) \cite{BerryRobbins1997}
    & Bose/Fermi statistics derived from a geometric construction of exchange and spin within a specific bundle framework
    & Particular fibre-bundle structure over configuration space; smooth, geometrically defined exchange operations compatible with spin representations; prescribed form of parallel transport / geometric phase
    \\
    \midrule

    Balachandran et al.\ (1990s) \cite{Balachandran1993}
    & Spin--statistics-type constraints from geometric and topological arguments
    & Locality; fixed spatial topology; specific global and local regularity conditions on fields or wavefunctions
    \\
\end{longtable}

\subsection{An Eighty-Year Investigation: Existing Intermediate Particle Exchange Statistics}

The long-standing interest in particle exchange statistics beyond the Bose--Fermi dichotomy is closely tied to the scope and limitations of existing no-go theorems. On the one hand, relativistic no-go results—most notably the spin--statistics theorem and its algebraic generalizations—rely on a set of structural assumptions such as locality, relativistic causality, and positivity \cite{1940Pauli,doplicher1969fields,doplicher1971local,doplicher1974local}. While these theorems rigorously exclude non-Bose--Fermi statistics under their respective premises.

On the other hand, attempts to establish no-go theorems likewise been shown to depend on additional assumptions whose status and generality are not always transparent \cite{girardeau1965permutation,flicker1967symmetrization,bros1966theoretical,Laidlaw1971,leinaas1977theory,BerryRobbins1997,Balachandran1993,balachandran1991classical,kaplan2013pauli,kuckert2005spin}. As a result, it has remained unclear to what extent these arguments genuinely rule out intermediate particle exchange statistics, or merely exclude certain classes of constructions.

Against this backdrop, a large body of work has instead pursued the direct construction of alternative statistical schemes. These efforts are motivated either by an explicit attempt to circumvent existing no-go results, or by the more cautious recognition that the precise domain of validity of such no-go theorems may itself be incomplete or assumption-dependent.

Since the early 1940s, the possibility of particle exchange statistics interpolating between Bose and Fermi has therefore been investigated from a wide range of perspectives. As summarized  in Table~\ref{tab:intermediate_statistics}, these efforts span more than eighty years and can be broadly grouped into several conceptually distinct approaches, differing both in their starting points and in their interpretation of what ``intermediate statistics'' is meant to represent.

\begin{table}[t]
\centering
\caption{Representative approaches to intermediate statistics developed over the past eight decades. 
The last category corresponds to effective models in which intermediate-statistical behavior emerges from specific dynamical interactions rather than a modification of the genuine exchange principle.}
\label{tab:intermediate_statistics}
\renewcommand{\arraystretch}{1.25}
\begin{tabular}{p{4.2cm} p{10.5cm}}
\toprule
\textbf{Approach} & \textbf{Representative works} \\
\midrule

Statistical--mechanical generalizations 
&
Gentile (1940)\cite{gentile1940itosservazioni};  Haldane (1991)\cite{haldane1991fractional}; Wu (1994) \cite{wu1994statistical}; Dai \& Xie (2004) \cite{dai2004gentile}
\\[4pt]

Quantum--mechanical approaches
&
Green (1953) parastatistics \cite{green1953generalized}; Ohnuki \& Kamefuchi (1982)\cite{ohnuki1982quantum,Ohnuki1982}; Biedenharn (1989)\cite{biedenharn1989quantum}; Macfarlane (1989)\cite{macfarlane1989q};\\
& Greenberg (1990, 1991) quon\cite{greenberg1991particles} and infinite statistics \cite{greenberg1990example}; Tichy \& Mølmer (2017) immannons \cite{tichy2017extending}; Wang \& Hazzard (2025) \cite{wang2025particle}
\\[4pt]

Symmetric--function approaches
&
Chaturvedi (1996) \cite{chaturvedi1997interpolations}; Schmidt \& Schnack (2002) \cite{schmidt2002partition}; Zhou--Chen--Dai (2022) \cite{zhou2022unified}
\\[4pt]

Dynamical and effective--model approaches
&
Calogero (1971) \cite{calogero1971solution}; Haldane \& Shastry (1988) \cite{haldane1988exact}; Polychronakos (1992) \cite{polychronakos2002generalized,polychronakos1992exchange}
\\

\bottomrule
\end{tabular}

\end{table}

\paragraph{Statistical-mechanical generalizations.}
One of the earliest routes was pursued at the level of statistical mechanics, beginning with Gentile’s proposal of finite occupancy statistics \cite{gentile1940itosservazioni}. In this line of work, the Bose and Fermi distributions are modified by altering the allowed occupation numbers or counting rules, without an explicit reference to an underlying microscopic exchange principle. Later developments include Haldane’s formulation of fractional exclusion statistics \cite{haldane1991fractional}, its thermodynamic elaboration by Wu \cite{wu1994statistical}. While these constructions yield well-defined distribution functions and thermodynamic relations, their connection to quantum statistics is indirect and missing.

\paragraph{Quantum-mechanical approaches.}
A second line of investigation proceeds within quantum mechanics itself by generalizing the algebraic structure of creation and annihilation operators. The earliest and most influential example is Green’s parastatistics \cite{green1953generalized}, systematically developed in the monograph by Ohnuki and Kamefuchi \cite{ohnuki1982quantum,Ohnuki1982}. Related ideas include infinite statistics and the quon algebra introduced by Greenberg \cite{greenberg1990example,greenberg1991particles}, as well as algebraic constructions based on deformed oscillator algebras and generalized commutation relations \cite{biedenharn1989quantum,macfarlane1989q}. More recent proposals, such as immannons \cite{tichy2017extending} and related constructions \cite{wang2025particle}, also fall into this category. These approaches provide explicit Hilbert-space realizations of generalized statistics, but typically enlarge the space of microscopic degrees of freedom of particles.

\paragraph{Symmetric-function and combinatorial approaches.}
A third perspective emphasizes the role of symmetric functions and combinatorial structures in the counting of many-body states. Representative examples include the work of Chaturvedi \cite{chaturvedi1997interpolations}, Schmidt and Schnack \cite{schmidt2002partition}, and more recent developments by Zhou, Chen, and Dai \cite{zhou2022unified}. In this framework, intermediate statistics emerge from generalized partition functions, often expressed in terms of symmetric polynomials. While mathematically transparent, these constructions again do not uniquely fix the underlying microscopic interpretation of particle exchanges.

\paragraph{Dynamical and effective-model approaches.}
A substantial class of models realizes intermediate-statistical behavior through explicit interactions or constraints in the Hamiltonian. Prototypical examples include the Calogero model and its generalizations \cite{calogero1971solution}, as well as later developments by Haldane, Shastry, and Polychronakos \cite{haldane1991fractional,polychronakos2002generalized,polychronakos1992exchange}. In these systems, spectra and thermodynamics exhibit features reminiscent of intermediate statistics; however, the statistical behavior arises from dynamics rather than from the genuine exchange principle. As such, these models are most naturally interpreted as effective descriptions rather than candidates for a genuine new  particle exchange statistics. These broader generalizations, however, lie outside the scope of this work.

\subsection{Summary}

Despite more than eighty years of intensive investigation, the question whether particle exchange statistics beyond the Bose--Fermi dichotomy can exist in three-dimensional quantum systems remains unresolved. On the one hand, existing no-go theorems are derived under specific structural assumptions, and it is in principle possible to construct alternative statistical schemes by relaxing or circumventing these premises. On the other hand, assessing the physical legitimacy of such constructions requires embedding them consistently within the very frameworks to which the no-go arguments apply, making it often unclear whether a proposed particle exchange statistics genuinely evades a given no-go theorem or merely violates one of its underlying assumptions.

As a result, the current literature lacks a unified and operational criterion that can decisively determine whether a given proposal represents a genuine new particle exchange statistics or an effective description relying on additional degrees of freedom, modified dynamics, or hidden structural assumptions. The absence of such a criterion has led to a proliferation of formally distinct constructions whose mutual relations and physical status remain difficult to compare.

\textbf{These considerations point to the pressing need for a clear, model-independent, and practically applicable diagnostic that can systematically evaluate candidate particle exchange statistics within a well-defined framework}. Establishing such a criterion, and clarifying the precise boundary between admissible and inadmissible extensions of particle exchange statistics in three dimensions, is the primary objective of the analysis presented in the following sections.

\section{Fundamental Premises of Our No-Go Theorem } \label{secpre}

Quantum particles are classified into distinct categories according to their collective behaviors, with bosons and fermions being the most familiar examples. 
Any system of identical quantum particles should satisfy the fundamental postulates of quantum mechanics together with the additional requirements that encode particle indistinguishability.

We find that a key to eliminating this ambiguity lies in considering intermediate exchange statistics in a manner that simultaneously enforces consistency between the many-body Hilbert-space structure of quantum mechanics and the statistical microstate counting of indistinguishable particles. However, this critical consideration is seldom incorporated in existing research.
In practice, proposals typically modify either the Hilbert-space structure \cite{green1953generalized,greenberg1990example,cattani1984general,ohnuki1982quantum,okayama1952generalization,tichy2017extending} or the combinatorial counting of microstates \cite{gentile1940itosservazioni,dai2004gentile,haldane1991fractional,wu1994statistical}, but rarely both in a single consistent framework. Yet, for  quantum particles (i.e., exchange statistics of identical and indistinguishable particles), these two aspects are inseparable: physically meaningful exchange statistics must admit both a well-defined many-body Hilbert space realization in quantum mechanics \emph{and} a consistent counting of microstate for indistinguishable particles in statistical mechanics. Whether any single construction can satisfy both requirements is often left implicit.

In this section, we present the premises of our no-go theorem, which are based on the consistency of the quantum statistics of indistinguishable particles and the many-body Hilbert space structure in quantum mechanics.

\subsection{Quantum statistics of indistinguishable particles} 
Identical particles are fundamentally indistinguishable: assigning labels to them has no observable physical consequence. In quantum mechanics, this principle requires that exchanging any two identical particles leaves all measurable probabilities invariant \cite{dirac1981principles}. 
At the level of state vectors, this means that particle permutations are implemented by a unitary representation of the symmetric group $S_N$ on the $N$-particle Hilbert space  in three dimensions \cite{messiah1964symmetrization}.
While this represents the standard consensus on quantum indistinguishability, our no-go theorem grounds its requirement of indistinguishability specifically in \emph{statistical mechanics}. We therefore emphasize that our framework primarily relies on indistinguishability as manifested in the statistical ensemble description, which we now proceed to elaborate.

In \textbf{statistical mechanics}, the system we consider is the one containing the indistinguishable particles, where the change of particle labels does not lead to a new microstate \cite{pathria2017statistical,reichl2016modern}. It means that we count microstates for a system consisting of indistinguishable particles. The exchange of the labels of the particle indices will not generate a new physical microstate. Therefore, for a system of non-interacting particles, each occupation-number configuration (i.e., the set of occupation numbers for all single-particle states) corresponds to a unique microstate. 

The occupation numbers of energy levels do not uniquely specify a microstate due to the degeneracy of those levels. However, this non-uniqueness is not fundamental; it can be resolved by switching to the occupation numbers of individual single-particle states, which then do correspond to a unique microstate.

By contrast, for distinguishable particles, even if the occupation numbers for each single-particle state are specified, one must also consider which specific particle occupies each state, since different assignments lead to physically distinct microstates. 
For $N$ particles placed in single-particle states, all permutations are counted as different physical realizations of the system. As a result, specifying occupation numbers alone does not uniquely determine the microstate for distinguishable particles.  

When maintaining particle indistinguishability, if the occupation numbers correspond to microstates with additional multiplicity, the introduction of extra information becomes necessary. These specific cases will be discussed individually with concrete examples.

\subsection{Many-body Hilbert Space in Quantum Mechanics}
\label{baseInd}
In  \textbf{quantum mechanics}, given an $m$-dimensional Hilbert space, the inner product is invariant under $U(m)$-transformation \cite{dirac1981principles}. Therefore, 
it is arbitrary to choose the basis of a single-particle Hilbert space. 
In particular, the structure of the multi-particle Hilbert space constructed from taking the tensor product of the single-particle Hilbert space must be independent of the choice of single-particle basis in three dimensions.

These premises encapsulate the principles of indistinguishability and single-particle-basis independence as the fundamental requirements on the many-body Hilbert-space structure of quantum mechanics and the statistical microstate counting of indistinguishable particles. They form the foundational principles that \emph{any} proposed type of quantum statistics must satisfy to remain compatible between quantum mechanics and statistical mechanics. The mathematical expression of the two premises will be shown in Sec.~\ref{sec:canonical}.

\section{Label Configuration Space and Permutation Symmetry}

In nonrelativistic quantum mechanics, the configuration of $N$ point
particles in $d$-dimensional Euclidean space is usually described
by points 
\[
(x_{1},\dots,x_{N})\in(\mathbb{R}^{d})^{N},
\]
where $x_{i}\in\mathbb{R}^{d}$ denotes the position of the $i$-th
particle. If the particles are \emph{distinguishable}, each index
$i$ carries physical meaning (e.g.\ different masses, charges, etc.),
and relabeling $i\leftrightarrow j$ generates physical measurable
effects. For \emph{identical particles}, however, the labels are fictitious:
any two configurations that differ only by a relabeling are physically
indistinguishable. In this section, we will understand the particle exchange statistics
from the topology aspect. 

\subsection{Label Configuration Space}

Consider $N$ point particles in $\mathbb{R}^{d}$. The \emph{label}
configuration space is $Q_{\mathrm{label}}=(\mathbb{R}^{d})^{N}$, whose
elements we denote $\mathbf{x}=(x_{1},\dots,x_{N})$ with $x_{i}\in\mathbb{R}^{d}$.
To define the exchanges without collision, we can remove the diagonal 

\begin{equation}
\Delta=\{(x_{1},\dots,x_{N})\mid x_{i}=x_{j}\ \text{for some }i\neq j\},
\end{equation}
and define the reduced labeled configuration space 
\begin{equation}
Q_{\mathrm{red}}=Q_{\mathrm{label}}\setminus\Delta.
\end{equation}
The points of $Q_{red}$ correspond to ordered $N$-tuples of distinct
points in $\mathbb{R}^{d}$.

Provided an order of the finite set of labels $\{1,2,\dots,N\}$,
any bijection of this set onto itself is a permutation, and the set
of all permutations forms the symmetric group $S_{N}$. Explicitly,
a permutation $\pi\in S_{N}$ acts on the label configurations by
\emph{relabeling} the particles: 
\begin{equation}
P_{\pi}:Q_{\mathrm{label}}\to Q_{\mathrm{label}},\qquad P_{\pi}(x_{1},\cdots,x_{N})=(x_{\pi(1)},\cdots,x_{\pi(N)}).
\label{eq:perm-action}
\end{equation}
Clearly, $P_{\pi}$ is a bijection, and the map $\pi\mapsto P_{\pi}$
is an injective group homomorphism $P_{\sigma}\circ P_{\pi}=P_{\sigma\circ\pi}$, with $P_{\mathrm{id}}=\mathrm{id}$ and $P_{\pi^{-1}}=P_{\pi}^{-1}$. We
remark that the fact that the relabelings form a group $S_{N}$ holds
at any spatial dimension. In the literature, such actions of $S_{N}$
on the label configurations are typically regarded as passive transformations,
corresponding to changes of labeling rather than changes of physical
configuration. 

\subsection{Configuration Space of Identical Particles and Fundamental Group}

Start from the reduced labeled configuration space $Q_{red}$ by removing
the coordinate coincidences, on which $S_{N}$ acts freely (i.e.\ no
non-trivial permutation leaves a point fixed, since $x_{i}\neq x_{j}$).
The physical configuration space for $N$ identical particles is then
$Q_{\mathrm{phy}}=Q_{red}/S_{N}$. The central topological object
is the fundamental group $\pi_{1}(Q_{\mathrm{phy}})$, the group of
homotopy classes of closed loops in $Q_{\mathrm{phy}}$. Physically,
a closed loop in $Q_{\mathrm{phy}}$ represents a process in which
the particle positions evolve in time and return to their initial setup, up to permutations, without any collisions. 

In quantum mechanics, the wave function is single-valued. However,
the physical configuration space $Q_{\mathrm{phy}}$ of identical
particles is often not simply connected with holes from excluding collisions
and noncontractible loops when exchanging particles. Thus, quantum
wave functions live on the universal cover $\tilde{Q}_{phy}$, and
the fundamental group $\pi_{1}(Q_{\mathrm{phy}})$ acts as covering
transformations. 

Every element $g\in\pi_{1}(Q_{\mathrm{phy}})$ defines a covering
transformation $\mathbf{x}\rightarrow g\cdot\mathbf{x}$. Then a quantum
state is specified by a unitary representation
\begin{equation}
\rho:\pi_{1}(Q_{\mathrm{phy}})\to U(\mathcal{H})
\end{equation}
that governs the transformation of wave functions under nontrivial
loops, which $U(\mathcal{H})$ is the unitary group acting on Hilbert
space $\mathcal{H}$. The choice of the representation $\rho$ encodes
the possible particle exchange statistics. 

In particular, the fundamental group $\pi_{1}(Q_{\mathrm{phy}})$
varies according to spatial dimensions. In $d\ge3$, $\pi_{1}(Q_{\mathrm{phy}})\cong S_{N}$.
In $d=2$, $\pi_{1}(Q_{\mathrm{phy}})\cong B_{N}$ is the braid group
and in $d=1$, $\pi_{1}(Q_{\mathrm{phy}})$ is trivial for point particles:
there is no nontrivial exchange loop. 

\section{Mathematical Background} 
\label{sec:math}

The key mathematical aspect of our no-go theorem lies in leveraging the representation theory of unitary and permutation groups. Using mathematical tools such as symmetric functions and integer partitions, we translate our two fundamental assumptions, proposed in Sec.~\ref{secpre}, into equivalent mathematical constraints. Then, by utilizing properties of Kostka numbers in combinatorics, we establish the no-go theorem.

This section provides a brief review of the mathematical background, including the symmetric group \cite{littlewood1977theory}, the integer partition \cite{zhou2018statistical}, the symmetric function \cite{macdonald1998symmetric}, and the Schur-Weyl duality \cite{weyl1946classical}.

\subsection{The Symmetric Group and the Integer Partition}
The symmetric group $S_N$, which is faithfully represented by the $N!$ distinct permutations of $N$ particles, plays a fundamental role in understanding identical-particle systems. 

Any representation $\rho$ of $S_N$ can be decomposed into a direct sum of its irreducible representations (irreps), each weighted by their multiplicity \cite{littlewood1977theory}. For example, the $6$-dimensional regular representation of $S_3$ decomposes into the direct sum of its irreducible representations, with multiplicities equal to their dimensions: namely, two one-dimensional irreps (each with multiplicity 1) and one two-dimensional irrep (with multiplicity 2).

The irreducible representations of the symmetric group $S_N$ are uniquely labeled by the integer partitions of $N$. A partition is a way of writing $N$ as a sum of positive integers \cite{zhou2018statistical}. For instance, $S_3$ has three irreducible representations, corresponding to the partitions $(3)$, $(2,1)$, and $(1,1,1)$. In general, the number of irreducible representations of $S_N$ is given by the partition function $P(N)$, which counts the number of unrestricted integer partitions of $N$. As examples, $P(4)=5$ and $P(5)=7$.

More explicitly, an integer partition of $N$, denoted by $(\lambda) = (\lambda_1, \lambda_2, \ldots, \lambda_{\ell(\lambda)})$, is a finite sequence of positive integers (called parts) arranged in non-increasing order: $\lambda_1 \geq \lambda_2 \geq \cdots \geq \lambda_{\ell(\lambda)} > 0$, such that $|\lambda|=\sum_{i=1}^{\ell(\lambda)} \lambda_i = N$. The number of parts, $\ell(\lambda)$, is called the length of the partition.

We order the integer partitions of $N$ by the following rules: $(\lambda)$ precedes $(\lambda)'$ if $\lambda_1 > \lambda_1'$, or if $\lambda_1 = \lambda_1'$ but $\lambda_2 > \lambda_2'$, and so on. We denote the $J$th partition as $(\lambda)_J$. For example, the partitions of $4$ in order are $(4)$, $(3,1)$, $(2,2)$, $(2,1,1)$, and $(1,1,1,1)$. For any integer $N$, the first partition in the natural ordering is always $(N)$, while the last is $(1^N)$, where the superscript $N$ indicates that $1$ appears $N$ times. We denote $\lambda_{J,i}$ as the $i$th part of the $J$th partition. 

For an irreducible representation of $S_N$ indexed by the $I$th integer partition $(\lambda)_I$, its dimension $f_{(\lambda)_I}$ is given by \cite{littlewood1977theory}
\begin{equation} \label{dimension}
f_{(\lambda)_I} = N! \frac{\prod_{1 \le i < j \le \ell(\lambda)_I} (\lambda_{I,i} - \lambda_{I,j} - i + j)}{\prod_{i=1}^{\ell(\lambda)_I} (\lambda_{I,i} + \ell(\lambda)_I - i)!}.
\end{equation}
For example, $f_{(3)} = 1$, $f_{(1,1,1)} = 1$, and $f_{(2,1)} = 2$.

\subsection{The Symmetric Polynomials}
A function $f(x_1, x_2, \ldots, x_n)$ is called \emph{symmetric} if it is invariant under any permutation of its variables by the symmetric group $S_n$ \cite{macdonald1998symmetric}. That is, for every $\sigma \in S_n$, we have 
\begin{equation}
    f(x_{\sigma(1)}, x_{\sigma(2)}, \ldots, x_{\sigma(n)}) = f(x_1, x_2, \ldots, x_n).
\end{equation}
For example, $f(x_1,x_2)=x_1+x_2$ is a symmetric function, and $g(x_1,x_2)=x_1^2+x_2$ is not a symmetric function.

Symmetric polynomials form an important class of symmetric functions. One important kind of symmetric polynomials is the \emph{monomial symmetric function} ($m$-function) $m_{(\lambda)}(x_1, x_2, \ldots, x_l)$. Each $m$-function associates with an integer partition $(\lambda) = (\lambda_1, \lambda_2, \ldots, \lambda_k)$ of $N$ and is defined as:
\begin{equation}
    m_{(\lambda)}(x_1, x_2, \ldots, x_l) = \sum_{\text{all permutations of } \lambda} x_{i_1}^{\lambda_1} x_{i_2}^{\lambda_2} \cdots x_{i_k}^{\lambda_k},
\end{equation}
where the summation extends over all distinct permutations of the indices $i_1, i_2, \ldots, i_k$ chosen from $\{1, 2, \ldots, l\}$ with $l \geq N$. The number of variables $l$ can be finite (with $l \geq N$) or infinite. There exists a one-to-one correspondence between partitions of $N$ and $m$-ffunctions of degree $N$. For example, the $m$-function with $3$ variables $x_1,x_2,x_3$ corresponding to integer partition $(\lambda)=(2,1)$ is 
\begin{equation}
    m_{(2,1)}(x_1, x_2, x_3) = x_1^2x_2+x_1^2x_3+x_2^2x_1+x_2^2x_3+x_3^2x_1+x_3^2x_2
\end{equation}

Another fundamental kind of symmetric polynomials is the \emph{Schur polynomials} ($s$-function) $s_{(\lambda)}(x_1, x_2, \ldots)$, which can be defined through irreducible characters of the symmetric group:
\begin{equation}
    s_{(\lambda)}(x_1, x_2, \ldots) = \frac{1}{N!} \sum_{(\mu)} \chi_{(\lambda)}^{(\mu)} \left( \prod_{j=1}^{N} j^{a_j} a_j! \right) \prod_{m=1}^k p_m(x)^{a_m},
\end{equation}
where the sum is over all partitions $(\mu)$ of $N$, $\chi_{(\lambda)}^{(\mu)}$ denotes the irreducible character of $S_N$ corresponding to partition $(\lambda)$ evaluated on the conjugacy class indexed by $(\mu)$ (the conjugacy class and the irreducible representation of $S_N$ are both indexed by integer partition of $N$ \cite{littlewood1977theory}), $a_m$ counts the number of parts $m$ in $(\mu)$, and $p_m(x) = \sum_i x_i^m$ is the power-sum symmetric function. In this work, we primarily consider systems with a finite number of variables. The cases involving infinitely many or even continuous variables can also be readily generalized within the $p_m(x)$. The $s$-function of degree $N$ are also in one-to-one correspondence with integer partitions of $N$.
For example, for $N=4$, the $s$-functions are 
\begin{align}
s_{(4)} & = \frac{1}{24}\left(p_1^4 + 6p_1^2p_2 + 3p_2^2 + 8p_1p_3 + 6p_4\right), \\
s_{(3,1)} & = \frac{1}{8}\left(p_1^4 + 2p_1^2p_2 - p_2^2 - 2p_4\right), \\
s_{(2,2)} & = \frac{1}{12}\left(p_1^4 + 3p_2^2 - 4p_1p_3\right), \\
s_{(2,1,1)} & = \frac{1}{8}\left(p_1^4 - 2p_1^2p_2 - p_2^2 + 2p_4\right), \\
s_{(1,1,1,1)} & = \frac{1}{24}\left(p_1^4 - 6p_1^2p_2 + 3p_2^2 + 8p_1p_3 - 6p_4\right),
\end{align}
where the arguments $(x_1,x_2,\ldots)$ are omitted for brevity.

The set of all symmetric polynomials of degree $N$ forms a vector space over the underlying field \cite{macdonald1998symmetric}. Both the $m$-function and $s$-function form complete bases for this vector space. Consequently, any symmetric polynomial of degree $N$ can be uniquely expressed as a linear combination of either basis. These two bases are related by the Kostka numbers $K_{\lambda\mu}$ \cite{macdonald1998symmetric}:
\begin{equation}\label{sm}
    s_{(\lambda)} = \sum_{\mu} K_{\lambda\mu} m_{(\mu)},
\end{equation}
where the sum ranges over all partitions $(\mu)$ of $N$, and $K_{\lambda\mu}$ counts the number of semistandard Young tableaux of shape $(\lambda)$ and content $(\mu)$ \cite{macdonald1998symmetric}. In the main text, for the sake of clarity, we use $I$ to label the $I$th integerpartition of $N$, then the Eq.~\eqref{sm} is written as $s_{I} = \sum_{I}^{P(N)} K^J_{I} m_{J}$. For $K^J_{I}$ the upper index, $J$, is the row index,and the lower index, $I$ is the column index. For example, the Kostka number for $N=3$, $N=4$, and $N=5$ read
\[
\begin{pmatrix}
1 & 0 & 0  \\
1 & 1 & 0  \\
1 & 2 & 1 \\
\end{pmatrix}, 
\begin{pmatrix}
1 & 0 & 0 & 0 & 0 \\
1 & 1 & 0 & 0 & 0 \\
1 & 1 & 1 & 0 & 0 \\
1 & 2 & 1 & 1 & 0 \\
1 & 3 & 2 & 3 & 1
\end{pmatrix}, 
\begin{pmatrix}
1 & 0 & 0 & 0 & 0&0&0 \\
1 & 1 & 0 & 0 & 0&0&0 \\
1 & 1 & 1 & 0 & 0&0&0 \\
1 & 2 & 1 & 1 & 0&0&0 \\
1 &2 & 2 & 1 & 1&0&0\\
1 &3 & 3 & 3 &2&1&0\\
1 &4&5 & 6 & 5&4&1
\end{pmatrix}.
\]

\subsection{Decomposition of Hilbert Space under $U(m)$ Representations}

Consider a system of $N$ identical particles. The full Hilbert space $V_N$ is given by the $N$-fold tensor product of the single-particle Hilbert space $V$:
\begin{equation}
    V_N = V^{\otimes N},
\end{equation}
where $V \simeq \mathbb{C}^m$ is the $m$-dimensional Hilbert space for a single particle. The space of pure physical states for a single particle corresponds to the projective Hilbert space $\mathcal{P}(V) \simeq \mathbb{C} P^{m-1}$ \cite{dirac1981principles}.

The Hilbert space $V_N$ naturally carries a representation of the unitary group $U(m)$, induced from the fundamental representation on the single-particle space $V$. Explicitly, the action of $g \in U(m)$ on a tensor product basis vector is given by:
\begin{equation}\label{uonv}
    g(e_{i_1} \otimes e_{i_2} \otimes \ldots \otimes e_{i_N}) = (g e_{i_1}) \otimes (g e_{i_2}) \otimes \ldots \otimes (g e_{i_N}),
\end{equation}
where $\{e_1, e_2, \ldots, e_m\}$ is an orthonormal basis of $V$.

The invariance of the many-particle Hilbert space under $U(m)$ transformations embodies the physical requirement that all observable quantities must be independent of the choice of orthonormal basis in $V$, as stipulated in the fundamental premises of Sec.~\ref{baseInd}. Consequently, the physical Hilbert space $\mathcal{H}_N$ of an $N$-particle system must decompose into irreducible representations (irreps) of $U(m)$. For a given $N$, these irreps are uniquely labeled by integer partitions $(\lambda)_I$ of $N$ with at most $m$ parts. The decomposition takes the form as
\begin{equation}\label{decompu}
    \mathcal{V}_N = \bigoplus_{I=1}^{P(N)} \left(V_{U(m)}^I\right)^{\oplus C^I} ,
\end{equation}
where $C^I$ are non-negative integers representing the multiplicity of each irreducible representation $V_{U(m)}^I$ in the decomposition.

The Schur-Weyl duality can be treated as an example of decomposition of the full $N$-particle Hilbert space  $V_N = V^{\otimes N}$ \cite{weyl1946classical}. Since the space $V_N$ also carries a natural representation of the symmetric group $S_N$. For a permutation $\sigma \in S_N$, its action on a basis vector in $V_N$, $e_{i_1} \otimes e_{i_2} \otimes \ldots \otimes e_{i_N}$, is defined by permuting the tensor factors:
\begin{align}\label{sonv}
    \sigma\left(e_{i_{1}}\otimes e_{i_{2}}\otimes\ldots\otimes e_{i_{N}}\right)=e_{\sigma_{i_{1}}}\otimes e_{\sigma_{i_{2}}}\otimes\ldots\otimes e_{\sigma_{i_{N}}}.
\end{align}
The actions of $S_N$ and $U(m)$ on $V_N$, Eqs.~\eqref{uonv} and ~\eqref{sonv}, commute with each other, $[S_N,U(m)]=0$. This commutativity allows for a simultaneous diagonalization of both group actions, which in turn reduces the full many-particle Hilbert space $V_N$ into its physically admissible subspaces \cite{weyl1946classical}. The most familiar examples are bosons and fermions, whose physical Hilbert spaces correspond to the fully symmetric subspace $\mathrm{Sym}^N(V)$ and the fully antisymmetric subspace $\wedge^N V$, respectively.

Mathematically, this structure is captured by Schur-Weyl duality \cite{weyl1946classical}, which provides a complete decomposition of the tensor product space $V^{\otimes N}$ under the joint action of $U(m)$ and $S_N$,
\begin{equation} \label{eq:sw}
   V_N = V^{\otimes N} = \bigoplus_I^{P(N)} \left( V^I_{U(m)} \otimes V_{S_N}^I \right)  \,,
\end{equation}
where the index $I$ runs over partitions of an integer $N$. $V_{U(m)}^I$ and $V_{S_N}^I$ are subspaces that carry irreducible representations (irreps) of $U(m)$ and $S_N$, respectively.
The multiplicity of the $I$th irreps of $S_N$ appearing in the decomposition is $\mathrm{dim}V_{U(m)}^I$. The multiplicity of the $I$th irreps of $U(m)$ appearing in the decomposition is $\mathrm{dim}V_{S_N}^I$. Thus, the total dimension of the $I$-component is $\mathrm{dim}V_{U(m)}^I\times \mathrm{dim}V_{S_N}^I$. For example, a system consists of $5$ particles with the dimension of the Hilbert space of a single particle \( V \) being $6$, the decomposition of the Hilbert space under the Schur-Weyl duality is given in Table~\ref{tab:hilbert_decomp}.

\begin{table}[t]
\centering
\caption{An example of decomposing a $5$-particle Hilbert space. The dimension of the single-particle Hilbert space is $6$. }
\label{tab:hilbert_decomp}
\begin{tabular}{cccccc}
\toprule
Index&subspace & $\mathrm{dim} \left(V^I_{U(m)} \otimes V_{S_N}^I\right)$  & $\mathrm{dim}V_{S_N}^I$ & $\mathrm{dim}V_{U(m)}^I$ \\
\midrule
$I=1$&\( V^{(5)}=V^{(5)}_{U(m)} \otimes V_{S_N}^{(5)} \)    & 252    & 1   & 252    \\
$I=2$&\( V^{(4,1)} \)    & 2016    & 4  & 504    \\
$I=3$&\( V^{(3,2)} \)    & 2100    & 5    & 420    \\
$I=4$&\( V^{(3,1^2)} \)    & 2016    & 6    & 336    \\
$I=5$&\( V^{(2^2,1)} \)    & 1050    & 5   & 210    \\
$I=6$&\( V^{(2,1^3)} \)    & 336    & 4     & 84    \\
$I=7$&\( V^{(1^5)} \)    & 6    & 1     & 6    \\
 &\( V^{\otimes 5} \)    & $7776=6^5$    & --    & --    \\
\bottomrule
\end{tabular}
\end{table}

\section{Canonical Partition Function of An Ideal $N$-Particle System} \label{sec:canonical}
The partition function can be formulated as  \cite{reichl2016modern,pathria2017statistical} 
\begin{equation}
  Z(\beta,N) \;=\; \mathrm{Tr}_{\mathcal H_N}\!\left(e^{-\beta H}\right)
  \;=\; \sum_{E} \Omega(E,N)\, e^{-\beta E}.
  \label{eq:Z}
\end{equation}
where $\beta = 1/(k_{\mathrm B} T)$ with $k_{\mathrm B}$ the Boltzmann constant and $T$ the temperature. 
$\Omega(E,N)$ denotes the number of microstates under given energy $E$ and particle number $N$. The first equality in Eq.~\eqref{eq:Z} is the quantum–mechanical definition of the partition function, and the second rewrites the trace as a statistical sum over energy eigenvalues.

\subsection{Canonical Partition Function from Quantum Mechanics }
The Hilbert space $\mathcal{H}_N$ of an ideal $N$-particle system decomposes into irreps of $U(m)$ as in Eq.~\eqref{decompu}. Substituting this decomposition into Eq.~\eqref{eq:Z} yields the $N$-particle partition function:
\begin{align}
Z(\beta,N) & = \mathrm{Tr}_{\mathcal{V}_N}( e^{-\beta \mathcal{H}_N}) \label{eq:Z1} \\
&= \sum_{I=1}^{P(N)} C^I \mathrm{Tr}_{V^I_{U(m)}}(e^{-\beta \mathcal{H}_N})\\
&\equiv \sum_{I=1}^{P(N)} C^I s_I(x_1,x_2,\cdots,x_m), \label{eq:Z2}
\end{align}
where we use the identity $\mathrm{Tr}_{V^I_{U(m)}}(e^{-\beta \mathcal{H}_N}) = s_I(x_1,x_2,\cdots,x_m)$ \cite{zhou2022unified}. Here, $\mathcal{H}_N$ is the system Hamiltonian, and $x_i = e^{-\beta \epsilon_i}$ with $\epsilon_i$ denoting the energy of a single-particle state denoted $i=1,\cdots,m$. Without loss of generality, we assume $m > N$.

As noted in Eq.~\eqref{decompu}, the coefficients $C^I$ represent the multiplicity of each irreducible representation $V_{U(m)}^I$ in the decomposition. Consequently, in Eq.~\eqref{eq:Z2}, these coefficients must be non-negative integers—an important constraint that forms a criterion in our no-go theorem. The expansion of canonical partition function in $s$-functions, Eq.~\eqref{eq:Z2}, can be functional as a \emph{mathematical formulation} of an ideal system of identical particles from the perspective of quantum mechanics, while the nonnegative coefficients $C^I$ specify the corresponding representations of $U(m)$ associated with basis independence. We can take bosons and fermions as eexamples;the canonical partition functions of bosons and fermions are $Z_\mathrm{boson}(\beta,N)=s_{(N)}(x_1,x_2,\cdots,x_m)$ and $Z_\mathrm{fermion}(\beta,N)=s_{(1^N)}(x_1,x_2,\cdots,x_m)$.

 As a remark, the physical Hilbert space $\mathcal{H}_N$ of an ideal $N$-particle system must be invariant under both $S_N$ and $U(m)$, decomposing into irreps of the form $V^I_{U(m)} \otimes V_{S_N}^I$ according to the Schur-Weyl duality, Eq.~\eqref{eq:sw}.  More precisely, if a particular irreps labeled by $I$ is included in the physical Hilbert space, then the whole block $V^I_{U(m)} \otimes V_{S_N}^I$ should be considered. Thus, the coefficient $C^I$ should not only be positive but also equals to $\dim V^{I}_{S_N}$ given in Eq.~\eqref{dimension} (since the Hamiltonian $\mathcal{H}_N$ does not act on the permutation degree of freedom); otherwise, $C^I = 0$. However, in the no-go theorem, we relax the strict symmetry requirement, imposing only that the $s$-function coefficients $C^I$ are non-negative. This preserves $U(m)$ invariance (at the cost of foregoing the strict $S_N$ invariance condition) while, as we will prove, providing sufficient criteria for our no-go theorem based on quantum statistical considerations.

\subsection{Canonical Partition Function from Operator Algebra}

From the perspective of quantum mechanics, Eq.~\eqref{eq:Z2} provides a general mathematical framework for ideal systems of identical particles. By selecting different irreducible representations of $S_N$ and $U(m)$, 
equivalently, by specifying a particular set of coefficients $C^I$, one can define distinct types of quantum systems.

In the second quantization language, identical particles are described by algebras of creation and annihilation operators. For instance, the canonical commutation relations $a_i a_j^\dagger \mp a_j^\dagger a_i = \delta_{ij}$, with the minus sign for bosons and the plus sign for fermions, encode the particle exchange statistics, giving rise to Bose-Einstein and Fermi-Dirac statistics, respectively. However, a fundamental open question remains: \textbf{how to systematically construct  operator algebras corresponding to \emph{arbitrary} irreducible representations of the symmetric group in the second quantization language.}

Conversely, given a specific operator algebra, one may ask whether it describes a consistent system of ideal identical particles. Currently, this question must be addressed on a case-by-case basis. Here, we establish a \emph{general criterion}:
\begin{tcolorbox}
    A system describes an ideal ensemble of identical particles \emph{if and only if} the canonical partition function derived from the many-particle Hamiltonian in the Fock space (Eq.~\eqref{eq:Z1}) matches the $s$-function expansion in Eq.~\eqref{eq:Z2} for the given single-particle energy levels.
\end{tcolorbox}
This criterion follows directly from our central result,  which establishes Eq.~\eqref{eq:Z2} as the universal form for canonical partition functions of all ideal systems of identical particles.

\subsection{Canonical Partition Function from Statistical Mechanics} \label{stapers}

\subsubsection{Conventional Treatment}
In statistical mechanics, the canonical partition function can be expressed in terms of occupancy numbers $(\lambda)$~\cite{reichl2016modern,pathria2017statistical},
\begin{equation}
Z(\beta,N)
=\sum_{E}\Omega(E,N)e^{-\beta E}
=\sum_{(\lambda_\epsilon)} \Omega^{(\lambda_\epsilon)}\exp\left(-\beta \sum_{\epsilon}\epsilon\lambda_\epsilon\right),
\label{z:conv}
\end{equation}
where $\lambda_\epsilon$ records the number of particles in the single-particle energy level $\epsilon$, and $(\lambda_\epsilon)$ denotes the collection of all such $\lambda_\epsilon$ giving the occupation-number pattern. The sum $\sum_{(\lambda_\epsilon)}$ runs over all distribution sets satisfying the constraint $\sum_{\epsilon}\lambda_\epsilon=N$. The factor $\Omega^{(\lambda_\epsilon)}$ is the statistical weight (the microstate count) appropriate to the distribution set $(\lambda_\epsilon)$. In the conventional treatment, the main difficulty in evaluating the canonical partition function lies in enumerating all admissible distribution sets. For example, consider a three-particle system with three single-particle energy levels $\epsilon_1,\epsilon_2,\epsilon_3$. The possible distribution sets include $(\lambda_\epsilon)=(0,1,2)$, $(0,2,1)$, $(1,0,2)$, $(1,2,0)$, $(2,1,0)$, $(2,0,1)$, etc. The sum $\sum_{(\lambda_\epsilon)}$ must simultaneously enforce particle-number conservation and account for permutations over all available energy levels. Moreover, when level degeneracies are present, the statistical weight $\Omega^{(\lambda_\epsilon)}$ must incorporate them. For bosons and fermions, for instance,
\begin{align}
\Omega_\mathrm{bosons}^{(\lambda_\epsilon)}=\prod_{\epsilon}\frac{(\lambda_\epsilon+g_\epsilon-1)!}{\lambda_\epsilon!(g_\epsilon-1)!},\label{omega_energy_b}\\
\Omega_\mathrm{fermions}^{(\lambda_\epsilon)}=\prod_{\epsilon}\frac{g_\epsilon!}{\lambda_\epsilon!(g_\epsilon-\lambda_\epsilon)!},\\
\label{omega_energy_f}
\end{align}
where $g_\epsilon$ is the degeneracy of level $\epsilon$. For distinguishable particle 
\begin{align}
\Omega_\mathrm{distinguishable}^{(\lambda_\epsilon)}=\prod_{\epsilon}N!\frac{g_\epsilon^{\lambda_\epsilon}}{\lambda_\epsilon!}.
\label{omega_energy_mb}
\end{align}

\subsubsection{Symmetric Polynomials Treatment}
In the main text, we express the canonical partition function in terms of the $m$-functions. While mathematically equivalent to the conventional treatment in Eq.~\eqref{z:conv}, this approach offers a significant conceptual and computational simplification.

The key insight is a shift from summing over specific occupation numbers of energy levels to summing over \emph{state occupation types}, which correspond to integer partitions of the total particle number $N$ \cite{chaturvedi1997microscopic,zhou2022unified}. For example, consider a three-particle system with three single-particle energy levels $\epsilon_1, \epsilon_2, \epsilon_3$. The conventional approach requires summing over numerous specific distributions like $(n_1,n_2,n_3) = (0,1,2)$, $(0,2,1)$, $(1,0,2)$, $(1,2,0)$, $(2,1,0)$, $(2,0,1)$, etc. Crucially, in terms of occupation-number pattern, all these distinct distributions belong to a single category: the partition $(\lambda) = (2,1)$ of the integer $3$.

This method offers a fundamental advantage: while occupation numbers of energy levels can be converted into occupation numbers of states, the former is complicated by level degeneracy. In contrast,\textbf{ our approach deals directly with the occupation of individual quantum states.}
Therefore, instead of tracking which specific states are occupied by how many particles, we only need to specify the occupation-number pattern—how many particles occupy a state, irrespective of the specific state. The $m$-function automatically incorporates all microscopic state configurations consistent with a given occupation-number pattern. That is, 
\begin{align}
Z(\beta, N) & = \sum_{(\lambda)} \Omega^{(\lambda)}\sum_{\text{distinct permutations of } \lambda} x_{i_1}^{\lambda_1} x_{i_2}^{\lambda_2} \cdots x_{i_k}^{\lambda_k} \notag \\
& = \sum_{(\lambda)} \Omega^{(\lambda)} m_{(\lambda)}(x_1, x_2, \cdots,x_m) \notag  \\
&\equiv \sum_{K=1}^{P(N)} \Omega^{K}\, m_{K}(x_1, x_2, \cdots,x_m) \label{eq:Zm}.
\end{align}

The coefficient $\Omega^{K}$ serves as an indicator function that specifies whether a given occupation-number pattern $(\lambda)_K$ is physically allowed: it takes the value $1$ if the occupation-number pattern is permitted, and $0$ otherwise. For instance, in the case of bosons, all occupation-number pattern are allowed, yielding the partition function $Z_{\text{boson}}(\beta,N) = \sum_{(\lambda)} m_{(\lambda)}(x_1,x_2,\cdots,x_m)$,
where $\Omega^{K} = 1$ for all partitions $(\lambda)_K$. In contrast, for fermions, only the occupation-number pattern $(1^N)$ (where each single-particle state is occupied by at most one particle) is permitted, resulting in the partition function $Z_{\text{fermion}}(\beta,N) = m_{(1^N)}(x_1,x_2,\cdots,x_m)$
with $\Omega^{P(N)}=\Omega^{(1^N)} = 1$ and $\Omega^{K} = 0$ for all other partitions $(\lambda)_K$. 

The expansion of canonical partition function in $m$-functions, Eq.~\eqref{eq:Zm}, can be functional as a \emph{mathematical formulation} of an ideal system of identical particles from the perspective of quantum statistical mechanics, while the binary choice, $\{0,1\}$, of coefficients $\Omega^K$ sspecifies the principles of indistinguishability. For example, for distinguishable particles, $\Omega^{K} = \frac{N!}{\prod_{j=1}^N \lambda_{K,j}!}>1$ if $N>1$.

\subsection{Summary on Physical Interpretation of Coefficients $C^I$ and $\Omega^{K}$}
The fundamental requirements of indistinguishability and basis independence are reflected in the structure of the canonical partition function through its expansions in $s$-functions and $m$-functions. Specifically, these principles impose stringent constraints on the coefficients $C^I$  and $\Omega^J$ appearing in these expansions, Eqs.~\eqref{eq:Z2} and ~\eqref{eq:Zm}, respectively.

\paragraph{The $s$-function side}
The coefficient $C^I$ in Eq.~\eqref{eq:Z2} denotes the multiplicity of the irrep block $\left( V^I_{U(m)} \otimes V_{S_N}^I \right)$ labeled by the partition $(\lambda)_I$ in Eq.~\eqref{eq:sw}. For the Hilbert space to consistently support simultaneous representations of both $S_N$ and $U(m)$, we relax the strict requirement and only demand that the coefficients $C^I$ be non-negative. This retains the criterion for $U(m)$ invariance at the cost of losing the specific criterion for $S_N$ invariance. Any violation of this non-negativity, i.e., $C^I < 0$, signals a physical inconsistency: the system becomes dependent on a specific basis choice, indicating a breakdown of $U(m)$ invariance.

\paragraph{The $m$-function side.} The coefficient $\Omega^{K}$ in Eq.~\eqref{eq:Zm} counts the number of microstates compatible with a given occupation-number pattern. For a system of indistinguishable particles where single-particle states are fundamentally distinguishable, the microstate is uniquely determined. Thus, one expects the $\Omega^{K}\in\{1,0\}$ for any configuration of state occupations. A deviation, $\Omega^{K} > 1$, signals a violation of a fundamental premise: either the indistinguishability of particles has been compromised, or the microstate cannot be determined only by the occupation-number pattern. 

An overview of violations of the standard forms of $C^I$ and $\Omega^K$ is presented in Table~\ref{tab:vioofc}. Although we do not provide a criterion for detecting violations of $S_N$ invariance in the Hilbert space, $\Omega^K$ serves as a crucial criterion for identifying breakdowns of indistinguishability. Thus, the standard forms of $C^I$ and $\Omega^K$ together constitute a complete no-go criterion.

\begin{table}[t]
\centering
\caption{Criteria and physical interpretation of violations for the coefficients $C^I$ and $\Omega^K$. A violation ($C^I < 0$) indicates a breakdown of basis independence, whereas the criterion does not assess $S_N$ invariance. The breakdown of $S_N$ invariance is a property properly assessed through the characters of the symmetric group $S_N$, rather than those of the unitary group $U(m)$ associated with s-functions.}
\label{tab:vioofc}
\begin{tabular}{p{2cm} p{2cm} p{10.5cm}}
\toprule
Coefficient & Violation Criterion & Physical Interpretation \\
\midrule
$C^I$ & $C^I < 0$ & Breakdown of basis independence ($U(m)$ invariance). \textit{Note: This is not a criterion for the breakdown of $S_N$ invariance.} \\
 $\Omega^K$ &  $\Omega^K > 1$ & Breakdown of particle indistinguishability \\
\bottomrule
\end{tabular}
\end{table}

\section{No-go Theorem: Constraints from Kostka Numbers on Partition Function}

\subsection{Kostka Numbers and the Partition Function}

The canonical partition function admits dual representations:
\begin{itemize}
    \item A linear combination of $s$-functions with coefficients $C^I$ (Eq.~\eqref{eq:Z2})
    \item A linear combination of $m$-functions with coefficients $\Omega^K$ (Eq.~\eqref{eq:Zm})
\end{itemize}

The physical interpretation of these coefficients imposes fundamental constraints: $\Omega^K \in \{0,1\}$ (binary selection of the number of microstates for given occupation-number pattern) and $C^I \in \mathbb{N}_0$ (non-negative multiplicity in the Hilbert space decomposition).

Crucially, the Kostka-number relation between the bases (Eq.~\eqref{sm}) enforces the consistency condition:
\begin{equation}\label{eq:OmegaK}
  \Omega^{J} = \sum_{I=1}^{P(N)} k^{J}_{I} C^{I}.
\end{equation}

When these physical and combinatorial constraints are applied simultaneously, the only solutions to Eq.~\eqref{eq:OmegaK} correspond precisely to bosonic and fermionic statistics, thus excluding any intermediate possibilities.

\subsection{Proof of the no-go theorem}
\textbf{Review of Kostka numbers.} The Kostka number $k^{J}_{I}$ is equal to the total number of semistandard Young tableaux of shape $(\lambda)_J$ and weight $(\lambda)_I$. The Kostka matrix exhibits the following properties\cite{littlewood1977theory,macdonald1998symmetric}:
First, it is a lower‐triangular matrix with all diagonal entries equal to 1, $k_I^I=1$ and $k_J^I\geq1$ when $I>J$. Second, first column consists entirely of ones, that is \(k^{I}_{J=1}=1\).

To prove the no-go theorem, we need the following lemma.

\textbf{Lemma}. There exists a solution of $\Omega^{I} = \sum_{J} k^{I}_{J} C^{J}$ with the constraint $C^I\in \mathbb N_0,\Omega^I\in \{0,1\}$ $\forall I$, if and only if the nonzero region of a given column in the Kostka matrix, say, the $L$th column, is filled entirely with $1$. In that case, $ C^{J=L}=1$ and $C^{J\neq L}=0$. for $K \ge L$ and $\Omega^{I}=0$ for other $I$.

\textbf{Proof of Lemma}. By using the properties of the Kostka number, Eq.~\eqref{eq:OmegaK} can be rewritten as  
\begin{equation}\label{eq:omegaK-sum}
  \Omega^{I} = \sum_{J} k^{I}_{J} C^{J}
  \;=\; C^{I} + \sum_{J<K} k^{I}_{J} C^{J}\, .
\end{equation}
Under the constraint $\Omega^{J} \le 1$,  must $C^J$ be $1$ or $0$.

If we let $C^{J=L}=1$, then Eq.~\eqref{eq:omegaK-sum} can be rewritten as
\begin{equation}\label{eq:omegaK-cases}
  \Omega^{I} =
  \begin{cases}
    C^{I} + \displaystyle\sum_{\substack{J<I\\ J\neq L}} k^{I}_{J} C^{J} + k^{I}_{L} C^{L}, & \text{if } I>L,\\[0.6ex]
    C^{I} + \displaystyle\sum_{J<I} k^{I}_{J} C^{J}, & \text{if } I\le L.
  \end{cases}
\end{equation}

As shown in Eq.~\eqref{eq:omegaK-cases}, to keep \(\Omega^{I} \le 1\), \(\ C^{J>I}=0\) and \(\ k_L^{I}=1\) should be hold for \(\ J>L\) and \(\ I>L\), respectively, or \(\Omega^{I}= C^{I} + \sum_{J<I,\, J\neq L} k^{I}_{J} C^{J} + k^{I}_{L} C^{L} \ge C^{I} + \sum_{J<I,\, J\neq L} k^{I}_{J} C^{J} + k^{I}_{L} \ge k^{I}_{L}\)  will be larger than $1$. \( C^{J}=0\) should also be held for \(\ J<L\), or \(\ C^{J=L}\) should be $0$, which contradicts with the \(\ C^{J=L}=1\). That is, if there exists a solution, it should be in the form \(\ C^{J=L}=1\) , \(C^{J\neq L}=0\) , \(\Omega^{I<L}=0\) and \(\Omega^{I \ge L}=1\). And this solution exists only when \(\ k_L^{I}=1\) holds for \(\ I>L\). That is, the nonzero region of the Lth column of Kostka number is filled entirely with $1$s.

It is straightforward to verify from the definition of Kostka numbers \cite{littlewood1977theory,macdonald1998symmetric} that only the first and last columns of the Kostka matrix are filled entirely with $1$s, corresponding to bosons and fermions. This property follows from the unique semi-standard fillings in these extreme cases. Therefore, we prove the no-go theorem.

\subsection{Examples of $C^I$ and $\Omega^K$} \label{sec_example}
We consider a system consisting of $N=5$ particles. The irreps of $S_5$ and $U(m)$ can be labeled by an integer partition of $5$.  For bosons and fermions, $C^{I}$ in Eq.~\eqref{eq:OmegaK} reads
\begin{equation}\label{bosec}
C_\mathrm{boson}=
\begin{pmatrix}
C^{1}=C^{(5)}=1 \\
C^{2}=C^{(4,1)}=0 \\
C^{3}=C^{(3,2)}=0 \\
C^{4}=C^{(3,1,1)}=0 \\
C^{5}=C^{(2,2,1)}=0 \\
C^{6}=C^{(2,1,1,1)}=0 \\
C^{7}=C^{(1,1,1,1,1)}=0 
\end{pmatrix}  \,\, \mathrm{and} \,\,
C_\mathrm{fermion}=
\begin{pmatrix}
C^{1}=0 \\
C^{2}=0 \\
C^{3}=0 \\
C^{4}=0 \\
C^{5}=0 \\
C^{6}=0 \\
C^{7}=1 
\end{pmatrix}.
\end{equation}

By using Eq.~\eqref{eq:OmegaK} with Kostka number \cite{macdonald1998symmetric}
\begin{equation}\label{eq:kostkaA1}
\bigl(k^{J}_{K}\bigr)=
\begin{pmatrix}
1 & 0 & 0 & 0 & 0 & 0 & 0\\
1 & 1 & 0 & 0 & 0 & 0 & 0\\
1 & 1 & 1 & 0 & 0 & 0 & 0\\
1 & 2 & 1 & 1 & 0 & 0 & 0\\
1 & 2 & 2 & 1 & 1 & 0 & 0\\
1 & 3 & 3 & 3 & 2 & 1 & 0\\
1 & 4 & 5 & 6 & 5 & 4 & 1
\end{pmatrix},
\end{equation}
one can obtain the corresponding $\Omega^{K}$. That is, for bosons and fermions, $\Omega^{K}$ in Eq.~\eqref{eq:OmegaK} reads
\begin{equation}\label{boseo}
\Omega_\mathrm{boson}=
\begin{pmatrix}
\Omega^{1}=\Omega^{(5)}=1 \\
\Omega^{2}=\Omega^{(4,1)}=1\\
\Omega^{3}=\Omega^{(3,2)}=1 \\
\Omega^{4}=\Omega^{(3,1,1)}=1 \\
\Omega^{5}=\Omega^{(2,2,1)}=1 \\
\Omega^{6}=\Omega^{(2,1,1,1)}=1 \\
\Omega^{7}=\Omega^{(1,1,1,1,1)}=1 
\end{pmatrix} \,\, \mathrm{and} \,\,
\Omega_\mathrm{fermion}=
\begin{pmatrix}
\Omega^{1}=0 \\
\Omega^{2}=0 \\
\Omega^{3}=0 \\
\Omega^{4}=0 \\
\Omega^{5}=0 \\
\Omega^{6}=0 \\
\Omega^{7}=1 
\end{pmatrix},
\end{equation}
where $\Omega_{boson}$ and $\Omega_{fermion}$  denotes the collections of $\Omega^{K}$ for bosons and fermions, respectively. 

Now, we consider paraparticle generalized through exchange symmetry \cite{green1953generalized,tichy2017extending,greenberg1990example,okayama1952generalization,chaturvedi1996canonical, wang2025particle}. Assume that the paraparticle is described by an irreps subspace labeled by the partition $(3,1,1)$ which gives a six-dimensional irreducible representation of $S_N$. That is, $C^{I}$ and the the corresponding $\Omega^{K}$ in Eq.~\eqref{eq:OmegaK} reads 
\begin{equation}\label{example}
C=
\begin{pmatrix}
C^{1}=0 \\
C^{2}=0 \\
C^{3}=0 \\
C^{4}=1 \\
C^{5}=0 \\
C^ {6}=0 \\
C^{7}=0 
\end{pmatrix} \,\xLeftrightarrow[\text{Kostka number}]\,
\Omega=
\begin{pmatrix}
\Omega^{1}=0 \\
\Omega^{2}=0 \\
\Omega^{3}=0 \\
\Omega^{4}=1 \\
\Omega^{5}=1 \\
\Omega^{6}=3 \\
\Omega^{7}=6 
\end{pmatrix},
\end{equation} 
where  $(3,1,1)$ is the fourth partition of $5$ and the corresponding $\Omega^{K}$ is calculated directly. $\Omega^{6}$ and $\Omega^{7}$ in Eq.~\eqref{example} mean that, for occupation-number pattern, $(\lambda)_6=(2,1,1,1)$ and $(\lambda)_7=(1,1,1,1,1)$, the number of distinct microstates are $3$ and $6$ rather than the required unique state. Therefore, it violates the principle of indistinguishability.

We can also consider intermediate statistics generalized through modifying the counting rules. Gentile statistics \cite{gentile1940itosservazioni}, generalized Gentile statistics (permit distinct maximum occupation numbers for different states) \cite{dai2004gentile}, and exclusive statistics (without topology, the fracture coefficient will be discussed in Sec.~\ref{fracture_discussion}) \cite{haldane1991fractional, wu1994statistical,fahssi2018combinatorics} all restrict the set of allowed occupation-number patterns with different constraints. Here, without the loss of generality, we consider a paraparticle with maximum occupation number $q=2$. In this case, the only admissible occupation-number patterns are $(\lambda)_7=(1,1,1,1,1)$, $(\lambda)_6=(2,1,1,1)$, and $(\lambda)_5=(2,2,1)$. That is, $\Omega^{K}$ and the corresponding $C^{I}$ in Eq.~\eqref{eq:OmegaK} reads:
\begin{equation} \label{example2}
\Omega=
\begin{pmatrix}
\Omega^{1}=0 \\
\Omega^{2}=0 \\
\Omega^{3}=0 \\
\Omega^{4}=0 \\
\Omega^{5}=1 \\
\Omega^{6}=1 \\
\Omega^{7}=1 
\end{pmatrix} \,\xLeftrightarrow[\text{Kostka number}]\,
C=
\begin{pmatrix}
C^{1}=0 \\
C^{2}=0 \\
C^{3}=0 \\
C^{4}=0 \\
C^{5}=1 \\
C^{6}=-1 \\
C^{7}=0 
\end{pmatrix}
\end{equation} 
$C^{6}=-1$ in Eq.~\eqref{example2} implies that the multiplicities of irreps labeled by \(\ (2,1,1,1)\) is negative. As a result, such particle violates the principle of indistinguishability or basis independence. Case by case analysis of specified generalized statistics will be given in Sec.~\ref{casebycase}.

\subsection{Examples from A Hilbert-Space View}\label{sec_example_hilbert}
Our no-go theorem establishes that any choice of an $N$-particle ``statistics sector'', i.e., a Hilbert subspace of $(\mathbb{C}^m)^{\otimes N}$ used to describe identical particles, which is neither the totally symmetric (bosonic) nor the totally antisymmetric (fermionic) subspace, necessarily contradicts the indistinguishability or basis independence. In this subsection we illustrate it by analyzing a few explicit examples at the level of Hilbert-space structure.

Let us consider a three-particle system whose single-particle Hilbert space $V$ is four-dimensional, $V \cong \mathbb{C}^4$. The total Hilbert space is then $V^{\otimes 3}$, which is $4^3 = 64$–dimensional, with a natural tensor-product basis $\{|i,j,k\rangle \mid 1 \le i,j,k \le 4\}$.

According to the Schur–Weyl duality decomposition, $V^{\otimes 3}$ splits into three irreducible $S_3$–sectors: a $20$–dimensional totally symmetric subspace (corresponding to the partition $(3)$), a $4$–dimensional totally antisymmetric subspace (corresponding to the partition $(1,1,1)$), and a $40$–dimensional subspace with mixed symmetry (corresponding to the partition $(2,1)$).

\subsubsection{$\bigoplus_{I\in \text{Sectors}} \left( V^I_{U(m)} \otimes V_{S_N}^I \right)$: 
Subspaces that Preserve Permutation and Unitary Invariance but Violate Indistinguishability}

If we describe the particles by restricting the physical Hilbert space to the $40$-dimensional mixed-symmetry sector $V^{(2,1)}_{U(4)} \otimes V^{(2,1)}_{S_3}$, then the canonical partition function of the system takes the form
\begin{align}
    Z \;=\; 2\,s_{(2,1)}.
\end{align}
By construction, this choice preserves both the single-particle $U(4)$ invariance and the $S_3$ permutation symmetry, because we keep the full isotypic component associated with the partition $(2,1)$.

However, in the basis of $m$-functions, we obtain
\begin{align}
    Z \;=\; 2\,m_{(2,1)} + 4\,m_{(1,1,1)}.
\end{align}
In particular, the occupation-number pattern of type $(1,1,1)$ (three particles occupying three distinct single-particle levels) appears with multiplicity $4$. In other words, there are four distinct Hilbert-space microstates corresponding to the same occupation-number pattern. This violates the requirement of indistinguishability that \emph{each physically distinct occupation-number pattern must be counted only once}.

\subsubsection{$\bigoplus_{I\in Sectors} \left( V^I_{U(m)}\right) $: Subspaces that Break the $S_N$ Invariance}
Within the $(2,1)$ isotypic component $V^{(2,1)}_{U(4)} \otimes V^{(2,1)}_{S_3}$
let us now restrict to a $20$–dimensional subspace obtained by fixing a nonzero vector 
$|\chi\rangle \in V^{(2,1)}_{S_3}$ and taking
\begin{align}
    \widetilde{\mathcal{V}}_{(2,1)} \;=\; V^{(2,1)}_{U(4)} \otimes \mathrm{span}\{|\chi\rangle\}
\;\cong\; V^{(2,1)}_{U(4)}.
\end{align}

This subspace still carries the irreducible representation of the unitary group $U(4)$, but it is \emph{not} invariant under the $S_3$–action. Writing out the basis vectors of $\widetilde{\mathcal{V}}_{(2,1)}$ explicitly:
\begin{equation}
\begin{aligned}
\Psi_1  &= |112\rangle + |112\rangle - |211\rangle - |211\rangle,\Psi_2  = |113\rangle + |113\rangle - |311\rangle - |311\rangle,\\
\Psi_3  &= |114\rangle + |114\rangle - |411\rangle - |411\rangle,\Psi_4  = |221\rangle + |221\rangle - |122\rangle - |122\rangle,\\
\Psi_5  &= |223\rangle + |223\rangle - |322\rangle - |322\rangle, \Psi_6 = |224\rangle + |224\rangle - |422\rangle - |422\rangle,\\
\Psi_7  &= |331\rangle + |331\rangle - |133\rangle - |133\rangle,\Psi_8  = |332\rangle + |332\rangle - |233\rangle - |233\rangle,\\
\Psi_9  &= |334\rangle + |334\rangle - |433\rangle - |433\rangle,\Psi_{10} = |441\rangle + |441\rangle - |144\rangle - |144\rangle,\\
\Psi_{11} &= |442\rangle + |442\rangle - |244\rangle - |244\rangle,\Psi_{12} = |443\rangle + |443\rangle - |344\rangle - |344\rangle,\\
\Psi_{13} &= |123\rangle + |213\rangle - |321\rangle - |312\rangle,\Psi_{14} = |132\rangle + |312\rangle - |231\rangle - |213\rangle,\\
\Psi_{15} &= |124\rangle + |214\rangle - |421\rangle - |412\rangle,\Psi_{16} = |142\rangle + |412\rangle - |241\rangle - |214\rangle,\\
\Psi_{17} &= |134\rangle + |314\rangle - |431\rangle - |413\rangle,\Psi_{18} = |143\rangle + |413\rangle - |341\rangle - |314\rangle,\\
\Psi_{19} &= |234\rangle + |324\rangle - |432\rangle - |423\rangle,\Psi_{20} = |243\rangle + |423\rangle - |342\rangle - |324\rangle.
\end{aligned}
\end{equation}
The canonical partition function in this case is 
\begin{align}
    Z=s_{(2,1)}=m_{(2,1)}+2m_{(1,1,1)},
\end{align}
where the coefficient of $m_{(1,1,1)}$ is $2$, indicating that two distinct Hilbert-space microstates correspond to the same occupation-number pattern. This violates the requirement of statistical indistinguishability that each occupation pattern be counted only once.

\paragraph{$\widetilde{\mathcal{V}}_{(2,1)}$ is not invariant under $S_3$} We now prove explicitly that $\widetilde{\mathcal{V}}_{(2,1)}$ is \emph{not} invariant under the action of $S_3$. It suffices to exhibit a vector $|\psi\rangle \in \widetilde{\mathcal{V}}_{(2,1)}$ and a permutation $P \in S_3$ such that $U(P)|\Psi\rangle \notin \widetilde{\mathcal{V}}_{(2,1)}$.

Consider the state vector
\begin{align}
    \Psi_{13} \;=\; |123\rangle + |213\rangle - |321\rangle - |312\rangle \;\in\; \widetilde{\mathcal{V}}_{(2,1)}.
\end{align}
Acting with the transposition $P_{12}$ (which exchanges the first two tensor factors), we obtain
\begin{align}
v=P_{12}\Psi_{13}= |213\rangle + |123\rangle - |231\rangle - |132\rangle.
\end{align}
By inspection of the definitions of $\Psi_1,\dots,\Psi_{20}$ above, we observe:

\begin{itemize}
  \item The basis kets $|123\rangle, |213\rangle, |132\rangle, |231\rangle$ appear \emph{only} in $\Psi_{13}$ and $\Psi_{14}$.
  \item None of the remaining vectors $\Psi_1,\dots,\Psi_{12},\Psi_{15},\dots,\Psi_{20}$ contain these kets in their expansion.
\end{itemize}

One can find that, there are no $\alpha,\beta\in\mathbb{C}$ such that $v = \alpha\Psi_{13} + \beta\Psi_{14}$,
and since no other $\Psi_i$ contain the kets $|123\rangle,|213\rangle,|132\rangle,|231\rangle$, we conclude that 
\begin{align}
    P_{12}\widetilde{\mathcal{V}}_{(2,1)} \neq \widetilde{\mathcal{V}}_{(2,1)}.
\end{align}
Therefore, $\widetilde{\mathcal{V}}_{(2,1)}$ is not invariant under the action of $P_{12}$, and hence not invariant under the full permutation group $S_3$.

\paragraph{An observable that distinguishes the two labelings}
We now define a perfectly legitimate observable on $\mathcal{H}$:
\begin{equation}
O \equiv\; |132\rangle\langle 132|.
\end{equation}
This operator projects onto the basis state $|132\rangle$; physically, it corresponds to asking:
\emph{``Are the three particles found in the configuration with particle~1 in level~1, particle~2 in level~3, and particle~3 in level~2?''}

One can find that
\begin{align}
\langle\Psi_{13}|O|\Psi_{13}\rangle
&= \langle\Psi_{13}|132\rangle\langle 132|\Psi_{13}\rangle =0.
\end{align}
Since $\Psi_{13}= |123\rangle + |213\rangle - |321\rangle - |312\rangle$
it has no $|132\rangle$ component. We can also evaluate the same observable in the exchanged state $v = P_{12}\Psi_{13}$.
We have $v= |213\rangle + |123\rangle - |231\rangle - |132\rangle$,
then, $\langle v|O|v\rangle= 1$. Thus,
\begin{align}
\langle P_{12}\Psi_{13}|O|P_{12}\Psi_{13}\rangle \neq \langle\Psi_{13}|O|\Psi_{13}\rangle.
\end{align}

It shows that once the state space is truncated to the $20$-dimensional subspace $W$ that is not invariant under $P_{12}$, the states $\Psi_{13}$ and $P_{12}\Psi_{13}$ are no longer physically equivalent:
there exist observables (such as $O$ above) for which their expectation values differ.

\subsubsection{$\bigoplus_{I\in \text{Sectors}} \left( V_{S_N}^I \right)$: Subspaces that Break $U(m)$ Invariance}

Now we consider the subspace $\widetilde{V}_{S_3}^{(2,1)}=span\{\phi_1,\phi_2\}$ with basis vectors
\begin{equation}\label{2dd}
    \begin{aligned}
\phi_1 &= |213\rangle - |312\rangle - |132\rangle + |123\rangle,\\
\phi_2 &= |213\rangle - |312\rangle - |321\rangle + |231\rangle.
\end{aligned}
\end{equation}
One can find that $\widetilde{V}_{S_3}^{(2,1)}$ carries the standard two-dimensional irrep of $S_3$ (the $(2,1)$ sector), and is invariant under the action of $S_3$. That is $U(P)\,\widetilde{V}_{S_3}^{(2,1)} \subset \widetilde{V}_{S_3}^{(2,1)}, \qquad \forall\,P\in S_3$.

\paragraph{$\widetilde{V}_{S_3}^{(2,1)}$ is not invariant under $U(4)$} 
The basis vectors $\phi_1$ and $\phi_2$ in Eq.~\eqref{2dd}, are constructed exclusively from the single-particle levels $\{|1\rangle,|2\rangle,|3\rangle\}$, with no contribution from $|4\rangle$. Consequently, any generic unitary transformation $U \in U(4)$ that mixes $|4\rangle$ with the other three levels will map $\widetilde{V}_{S_3}^{(2,1)}$ outside its original space, demonstrating the lack of $U(4)$ invariance.

\paragraph{Canonical partition function on $\widetilde{V}_{S_3}^{(2,1)}$}
When the system is restricted to $\widetilde{V}_{S_3}^{(2,1)}$, the canonical partition function becomes:
\begin{align}
    Z = 2\,e^{-\beta(\epsilon_1+\epsilon_2+\epsilon_3)},
\end{align}
This expression cannot be formulated as either an $s$-function or $m$-function expansion in the variables $\{x_1,x_2,x_3,x_4\}$, where $x_i = e^{-\beta\epsilon_i}$.

Beyond this particular case, Sec.~\ref{wang_model} examines additional examples of Hilbert spaces that break $U(m)$ invariance, yet still allow the canonical partition function to be expressed in both the $s$-function and $m$-function bases.

\subsubsection{Hilbert Space with A Maximum Occupation Number Constraint}

We continue with the totally symmetric $20$-dimensional subspace $\mathrm{Sym}^3(V)$ of the three-particle Hilbert space, where $V \cong \mathbb{C}^4$. We now impose a maximum occupation number constraint $q=2$, i.e., we forbid any configuration in which all three particles occupy the same single-particle level. At the level of basis states this amounts to removing the symmetric states of the form $|i,i,i\rangle$ ($i=1,\dots,4$), leaving a $16$-dimensional subspace
$W_{q=2} \subset \mathrm{Sym}^3(V).$
Every vector in $\mathrm{Sym}^3(V)$ is invariant under the action of $S_3$, and hence the same is true for its subspace $W_{q=2}$. Equivalently, $W_{q=2}$ is a direct sum of $16$ copies of the trivial one-dimensional representation of $S_3$, so the permutation symmetry is fully preserved.

By contrast, $W_{q=2}$ is not invariant under the full single-particle unitary group $U(4)$. A generic unitary transformation $U\in U(4)$ that mixes different single-particle levels will map an allowed state (with at most double occupancy) to a superposition that contains components with triple occupancy, and hence lies outside $W_{q=2}$. Thus the induced action $U^{\otimes 3}$ does not leave $W_{q=2}$ invariant for generic $U$, and no representation of the full $U(4)$ can be realized on this constrained subspace.

\paragraph{Canonical partition function on $W_{q=2}$}
In this constrained subspace, the canonical partition function can be written in terms of symmetric polynomials as
\begin{align}
    Z_{W_{q=2}}
    &= m_{(2,1)} + m_{(1,1,1)}
      \nonumber\\[4pt]
    &= s_{(2,1)} - s_{(1,1,1)}.
\end{align}
The $s$-function expansion thus involves a \emph{negative} coefficient. This signals that the corresponding partition function cannot arise from any genuinely $U(4)$-invariant free-particle Hilbert space: such a Hilbert space would necessarily yield a Schur-positive combination of $s$-functions. In other words, the maximum-occupation-number constraint explicitly breaks single-particle-basis independence and hence $U(4)$ invariance, which we will show that the observable will be depend on the choice of basis of single-particle Hibert space in Sec.~\ref{casebycase}. A summary is given in Table~\ref{tbl_hilbert_space}.

\begin{table}[t]
\centering
\caption{Possible choices of the system Hilbert space and their properties. Sectors denote the choice of subspaces. Specified examples will be discussed in Sec.~\ref{casebycase}.
}
\label{tbl_hilbert_space}
\begin{tabular}{ccccc}
\toprule
\multicolumn{1}{c}{\shortstack{Types}} & \multicolumn{3}{c}{\shortstack{Canonical \\partition function}} & \multicolumn{1}{c}{\shortstack{Physical\\Intepretation}} \\
\cmidrule(lr){2-4}
\shortstack{} & \shortstack{Admits $m$- and \\ $s$-function \\ expansion} & \shortstack{Standard $C^I$ \\ and $U(m)$ \\ invariance}& \shortstack{ Standard $\Omega^I$}  &    \\
\midrule
$\mathrm{Sym}^N(V)$ & $\checkmark$ & $\checkmark$ & $\checkmark$  & Allowed \\
$\wedge^N V$ & $\checkmark$ & $\checkmark$ & $\checkmark$  & Allowed \\
\midrule
$V^{\otimes N}$ & $\checkmark$ & $\checkmark$ & $\times$ & \shortstack{ Distinguashable }  \\
\cmidrule(lr){5-5}
$\displaystyle \bigoplus_{I\in \mathrm{Sectors}} \left( V^I_{U(m)} \otimes V_{S_N}^I \right) $ &$\checkmark$ & $\checkmark$ & $\times$ &  \shortstack{ Distinguashable }\\
\cmidrule(lr){5-5}
$\displaystyle \bigoplus_{I\in \mathrm{Sectors}} \left( V^I_{U(m)}\right) $ &  $\checkmark$ & $\checkmark$ & $\times$ & \shortstack{ Distinguashable }\\
\cmidrule(lr){5-5}
$\displaystyle \bigoplus_{I\in \mathrm{Sectors}} \left( V_{S_N}^I \right) $ & \shortstack{To be\\ undetermined}  &  $\times$ & $\times$  & \shortstack{Base\\ dependent and \\ Distinguashable}\\
\cmidrule(lr){5-5}
\shortstack{$\mathrm{Sym}^N(V)$\\with maximum \\ occupation \\number} & $\checkmark$ &  $\times$ & $\checkmark$  & \shortstack{Base\\ dependent}\\
\cmidrule(lr){5-5}

\bottomrule
\end{tabular}
\end{table}

\section{Generalized Statistics Violating the No-go Theorem}\label{casebycase}

This section examines various historical proposals for generalized quantum statistics. We apply our no-go theorem as a direct criterion to identify the specific points at which these intermediate statistics come into conflict with the fundamental principle of indistinguishability and basis independence in quantum mechanics. 

\subsection{Gentile Statistics}
Gentile statistics generalizes the Pauli exclusion principle by allowing the maximum occupation number of each quantum state to be an arbitrary integer $q$. This formulation recovers Bose-Einstein statistics when $q \to \infty$ and Fermi-Dirac statistics when $q = 1$. While several second quantization schemes have been proposed to implement this statistics, we will demonstrate that it fails as a quantum ideal identical-particle system.

\subsubsection{Canonical Partition Function} 

We now demonstrate how Gentile statistics \cite{gentile1940itosservazioni} violates the assumptions of our no-go theorem when treated as an ideal identical-particle system.
We consider a $4$ Gentile particles, with maximum occupation numbers $q$.

Due to the maximum occupation number constraint in Gentile statistics, some occupation-number patterns  allowed for bosons are forbidden. More specifically, the allowed occupation-number patterns  correspond to those partitions $I=(\lambda_1,\lambda_2,\cdots)$ with $\lambda_1\leq q$. Thus, we can write down the canonical partition function \cite{zhou2022unified}
\begin{equation} \label{eq:ZqGentile}
Z_{\text{Gentile,}q}(\beta,N) = \sum_{I,\lambda_{1}\leq q, |\lambda_i|=N} m_{I}(x_1,x_2,\dots).
\end{equation}
For example, with $q=2$ and $3$, given a $4$-particle Gentile system, the canonical partition functions are:
\begin{align}\label{eq:ZqGentile_m}
Z_{\text{Gentile,}q=2} &= m_{(2,2)} + m_{(2,1,1)} + m_{(1,1,1,1)},  \\
Z_{\text{Gentile,}q=3} &= m_{(3,1)} + m_{(2,2)} + m_{(2,1,1)} + m_{(1,1,1,1)},
\end{align}    
while the argument $(x_1,x_2,\dots)$ is omitted for brevity. 
By using the Eq.~\eqref{eq:OmegaK}, we obtain the canonical partition function expressed by $s$-function,
\begin{align} 
Z_{\text{Gentile,}q=2} &= s_{(2,2)} - s_{(1,1,1,1)}, \label{eq:ZqGentile_s}\\
Z_{\text{Gentile,}q=3} &= s_{(3,1)} - s_{(2,1,1)} + s_{(1,1,1,1)} .\label{eq:ZqGentile_s1}
\end{align}
In both cases, some coefficients are negative, indicating that no ideal quantum system corresponds to Gentile statistics with $q = 2$ or $q = 3$. We will show that Gentile statistics is basis dependent in Sec.~\ref{gentile_bain}  

\subsubsection{Counting Microstates: Statistical Indistinguashable Particles} One can check the statistics by enumerating all possible microstates directly. Consider a $4$-particle system with total energy $E=10$ and single-particle energy levels given by positive integers $1,2,\cdots$. For a system of bosons, the occupation-number pattern correspond to all possible ways of distributing the total energy $10$ among the four particles. For instance, the energy distribution \([7,1,1,1]\) means one particle occupies the state with energy $7$ and three particles occupy the state with energy $1$, which belongs to occupation-number pattern $(3,1)$ representing $3$ particles occupying the same state and $1$ particle occupying a different one. The complete list of such energy distribution is:
\begin{equation} \label{the_9_ed}
    \begin{aligned}
&[7,1,1,1],\ [3,3,3,1],\ [4,2,2,2],\ [4,4,1,1],\ [3,3,2,2],\\ 
&[5,3,1,1],\ [5,2,2,1],\ [6,2,1,1],\ [4,3,2,1]. 
\end{aligned}
\end{equation}

Here, one should not confuse the notation of energy distribution here with that of occupation-number pattern. 
Since the particles are indistinguishable, each occupation-number pattern corresponds to one microstate. Hence, for bosons, the total number of microstates is $9$. We can also obtain the number of microstates from the canonical partition function, accordingly, which can be calculated from the canonical partition function by using the inverse Z-transforms \cite{zhou2018statistical}. This is consistent with the bosonic partition function expressed in terms of $s$-functions and $m$-functions $
Z_{\text{Boson}} = s_{(4)}= m_{(4)} + m_{(3,1)} + m_{(2,2)} + m_{(2,1,1)}+ m_{(1,1,1,1)}$.

For the Gentile system with maximum occupation number $q = 2$, there are $6$ allowed microstates
\begin{align}
 [4,4,1,1],\ [3,3,2,2],\ [5,3,1,1],\ [5,2,2,1],\ [6,2,1,1],\ [4,3,2,1],
\end{align}
 where, compared with boson case, energy distributions such as $[7,1,1,1],\ [3,3,3,1],\ [4,2,2,2]$ are eliminated. Other cases can be found in Table~\ref{tbl_Gentile}.

The coefficients in Eq.~\eqref{eq:ZqGentile} are all $0$ or $1$, thus, the Gentile statistics is consistence with quantum statistical mechanics as iindistinguishableparticles.

\begin{table}[t]
\centering
\caption{Microstate counting for Gentile with $q=2,3,4$. Here we assume the single-particle states with energies: $1,2,3,\cdots$. We consider a $4$-particle system.}
\label{tbl_Gentile}
\begin{tabular}{ccccccc}
\toprule
\multicolumn{2}{c}{} & \multicolumn{5}{c}{Number of Microstates} \\
\cmidrule(lr){3-7}
\shortstack{Energy \\ Distributions} & \shortstack{Occupation \\ Type} & 
\shortstack{bosons} &\shortstack{fermions} &\shortstack{$q=2$} & \shortstack{ $q=3$} & \shortstack{ $q=4$}  \\
\midrule
$[7,1,1,1]$ & $(3,1)$ & $1$ & $0$ & $0$ & $1$ & $1$  \\
$[3,3,3,1]$ & $(3,1)$ & $1$ & $0$ & $0$ & $1$ & $1$ \\
$[4,2,2,2]$ & $(3,1)$ & $1$ & $0$ & $0$ & $1$ & $1$ \\
$[4,4,1,1]$ & $(2,2)$ & $1$ & $0$ & $1$ & $1$ & $1$\\
$[3,3,2,2]$ & $(2,2)$ & $1$ & $0$ & $1$ & $1$ & $1$\\
$[5,3,1,1]$ & $(2,1,1)$ & $1$ & $0$ & $1$ & $1$ & $1$\\
$[5,2,2,1]$ & $(2,1,1)$ & $1$ & $0$ & $1$ & $1$ & $1$ \\
$[6,2,1,1]$ & $(2,1,1)$ & $1$ & $0$ & $1$ & $1$ & $1$ \\
$[4,3,2,1]$ & $(1,1,1,1)$ & $1$ & $1$ & $1$ & $1$ & $1$ \\
\midrule
\shortstack{Total Number of \\ Microstates} & &9&1 & 6 & 9 & 9 \\
\bottomrule
\end{tabular}
\end{table}

\subsubsection{Examining Hilbert Space: Violation of  Basis Independence} \label{gentile_bain}

Our analysis ddemonstrates the incompatibility of Gentile statistics with fundamental quantum mechanics: the violation of $U(m)$ invariance. This is evident from the $s$-function expansion of the Gentile partition function in Eqs.~\eqref{eq:ZqGentile_s} and \eqref{eq:ZqGentile_s1}, where we find that the coefficients are not all non-negative. As highlighted by Greenberg \cite{greenberg1993quons}, the inconsistency can be illustrated by considering a simple example with two single-particle states $\lvert 1\rangle$ and $\lvert 2\rangle$, with maximum occupation number $q=2$. In this basis, the three-particle state $\lvert 1\rangle\lvert 1\rangle\lvert 2\rangle$ is allowed, while the state $\lvert 1\rangle\lvert 1\rangle\lvert 1\rangle$ is forbidden due to the occupation constraint. Now consider a transformation to a new basis:
\begin{align}
    \lvert +\rangle = \frac{1}{\sqrt{2}}(\lvert 1\rangle + \lvert 2\rangle), \quad 
\lvert -\rangle = \frac{1}{\sqrt{2}}(\lvert 1\rangle - \lvert 2\rangle).
\end{align}
When we express the originally allowed state $\lvert 1\rangle\lvert 1\rangle\lvert 2\rangle$ in this new basis, the expansion necessarily contains terms like $\lvert +\rangle\lvert +\rangle\lvert +\rangle$. However, such terms represent states where three particles occupy the same single-particle state $\lvert +\rangle$, which violates the maximum occupation number rule in the new basis. Similarly, other observables such as energy spectrum of the system also depend on the choice of basis in the single-particle Hilbert space, which we will show in Sec.~\ref{section:quon}.

By contrast, the maximum occupation number constraint of a fermionicsystem is preserved under basis transformations because the Hilbert space is invariant under unitary transformations.
The essential distinction is that the Hilbert space, constructed from the fully antisymmetrized $V_N$, transforms properly under the unitary group $U(m)$, ensuring that the antisymmetric property and the Pauli exclusion principle hold in any basis.

\subsection{Green's parastatistics}

Unlike Gentile statistics, Green's parastatistics provides a quantum mechanically consistent framework, characterized by its invariance under the unitary group $U(m)$ in the Hilbert space. However, we will demonstrate that its fundamental limitation emerges at the statistical level: the occupation-number pattern in parastatistics correspond to multiple distinguishable microstates, thereby failing to represent a genuine indistinguishable particle theory.

\subsubsection{Canonical Partition Function}
We also consider a system of $4$ parabosons or parafermions with order parameters $q = 2$, $3$, and $4$. The canonical partition functions for these systems are elegantly expressed in terms of $s$-functions \cite{chaturvedi1996canonical,zhou2022unified}, and these expressions are consistent with the results obtained from the commutation relations of Green's definition of creation and annihilation operators \cite{stoilova2020partition}. We therefore omit the derivation of the canonical partition function from the creation and annihilation operator formalism.

\noindent\textbf{paraboson systems:}
\begin{equation} \label{pb}
 \begin{aligned}
Z^{\text{PB}}_{q=2} & = s_{(4)} + s_{(3,1)} + s_{(2,2)}, \\[1mm]
Z^{\text{PB}}_{q=3} & = s_{(4)} + s_{(3,1)} + s_{(2,2)} + s_{(2,1,1)}, \\[1mm]
Z^{\text{PB}}_{q=4} & = s_{(4)} + s_{(3,1)} + s_{(2,2)} + s_{(2,1,1)} + s_{(1,1,1,1)}.
\end{aligned}
\end{equation}

\noindent\textbf{parafermion systems:}

\begin{equation}\label{pf}
\begin{aligned}
Z^{\text{PF}}_{q=2} & = s_{(1,1,1,1)} + s_{(2,1,1)} + s_{(2,2)}, \\[1mm]
Z^{\text{PF}}_{q=3} & = s_{(1,1,1,1)} + s_{(2,1,1)} + s_{(2,2)} + s_{(3,1)}, \\[1mm]
Z^{\text{PF}}_{q=4} & = s_{(4)} + s_{(3,1)} + s_{(2,2)} + s_{(2,1,1)} + s_{(1,1,1,1)}.
\end{aligned}
\end{equation}

By expressing the partition functions in the $m$-function basis, we see the microstate counting violates indistinguishability.

\textbf{paraboson systems ($m$-function basis):}
\begin{equation}
\begin{aligned}
Z^{\text{PB}}_{q=2} & = m_{(4)} + 2m_{(3,1)} + 3m_{(2,2)} + 4m_{(2,1,1)} + 6m_{(1,1,1,1)}, \label{pb2} \\[1mm]
Z^{\text{PB}}_{q=3} & = m_{(4)} + 2m_{(3,1)} + 3m_{(2,2)} + 5m_{(2,1,1)} + 9m_{(1,1,1,1)}, \\[1mm]
Z^{\text{PB}}_{q=4} & = m_{(4)} + 2m_{(3,1)} + 3m_{(2,2)} + 5m_{(2,1,1)} + 10m_{(1,1,1,1)}.
\end{aligned}
\end{equation}

\noindent\textbf{parafermion systems ($m$-function basis):}
\begin{equation}
\begin{aligned}
Z^{\text{PF}}_{q=2} & = 0m_{(4)} + 0m_{(3,1)} + m_{(2,2)} + 2m_{(2,1,1)} + 6m_{(1,1,1,1)}, \\[1mm]
Z^{\text{PF}}_{q=3} & = 0m_{(4)} + m_{(3,1)} + 2m_{(2,2)} + 4m_{(2,1,1)} + 9m_{(1,1,1,1)}, \\[1mm]
Z^{\text{PF}}_{q=4} & = m_{(4)} + 2m_{(3,1)} + 3m_{(2,2)} + 5m_{(2,1,1)} + 10m_{(1,1,1,1)}.
\end{aligned}
\end{equation}

Notably, when $q = 4$ (equal to the number of particles), paraboson and parafermion become the same.  

\subsubsection{Counting Microstates: Violation of Indistinguashablity}
To gain physical insight, we compute the number of microstates at fixed total energy \( E = 10 \), assuming a single-particle energy spectrum of positive integers by using the method of calculating the number of microstates from the canonical partition function \cite{zhou2018statistical}:
\begin{itemize}
\item parabosons: 30, 36, 37 microstates for \( q = 2, 3, 4 \), respectively;
\item parafermions: 14, 28 microstates for \( q = 2, 3 \), respectively.
\end{itemize}
Then, we count the microstates directly. Truly indistinguishable bosons yield only $9$ distinct energy partitions of $10$ among $4$ particles, see Eq.~\eqref{the_9_ed}.

For paraboson systems with $q=2$, the canonical partition function expressed in terms of monomial symmetric functions reveals that the occupation-number pattern $(3,1)$ - indicating three particles occupying one state and one particle occupying another state - should be counted twice, as evidenced by the coefficient $2$ for $m_{(3,1)}$ in the partition function, Eq.~\eqref{pb2}. Similarly, occupation-number patterns  of types $(2,2)$, $(2,1,1)$, and $(1,1,1,1)$ should be counted $3$, $4$, and $6$ times respectively.

Applying this multiplicity rule to the $9$ energy distributions listed above, we obtain a total of $30$ microstates, which matches the result calculated from the canonical partition function. The detailed calculation is summarized in the Table.~\ref{paracount}.

\begin{table}[h]
\centering
\caption{The numbers of microstates for parabosons/fermions with $q=2,3,4$. We assume the single-particle states with energies: $1,2,3,\cdots$. We consider a $4$-particle system while fixing the total energy to be $10$. }
\label{paracount}
\begin{tabular}{ccccccccc}
\toprule
\multicolumn{2}{c}{} & \multicolumn{7}{c}{Number of microstates} \\
\cmidrule(lr){3-9}
\shortstack{Energy \\ distri-\\butions} & \shortstack{Occup-\\ation \\ Type} & 
\shortstack{bosons} &\shortstack{fermions} &\shortstack{para-\\bosons \\ $q=2$} & \shortstack{para-\\bosons \\ $q=3$} & \shortstack{para-\\bosons \\ $q=4$} & \shortstack{para-\\fermions \\ $q=2$} & \shortstack{para-\\fermions \\ $q=3$} \\
\midrule
$[7,1,1,1]$ & $(3,1)$ & $1$ & $0$ & $2$ & $2$ & $2$& $0$& $1$ \\
$[3,3,3,1]$ & $(3,1)$ & $1$ & $0$ & $2$ & $2$ & $2$& $0$& $1$ \\
$[4,2,2,2]$ & $(3,1)$ & $1$ & $0$ & $2$ & $2$ & $2$ & $0$& $1$\\
$[4,4,1,1]$ & $(2,2)$ & $1$ & $0$ & $3$ & $3$ & $3$ & $1$& $2$\\
$[3,3,2,2]$ & $(2,2)$ & $1$ & $0$ & $3$ & $3$ & $3$& $1$& $2$ \\
$[5,3,1,1]$ & $(2,1,1)$ & $1$ & $0$ & $4$ & $5$ & $5$ & $2$& $4$\\
$[5,2,2,1]$ & $(2,1,1)$ & $1$ & $0$ & $4$ & $5$ & $5$& $2$& $4$ \\
$[6,2,1,1]$ & $(2,1,1)$ & $1$ & $0$ & $4$ & $5$ & $5$& $2$& $4$ \\
$[4,3,2,1]$ & $(1,1,1,1)$ & $1$ & $1$ & $6$ & $9$ & $10$& $6$& $9$ \\
\midrule
\shortstack{Total \\ number of \\ microstates} & &9&1 & 30 & 36 & 37 & 14& 28 \\
\bottomrule
\end{tabular}
\end{table}

\paragraph{Summary on Green's parastatistics}

The integer coefficients greater than $1$ explicitly demonstrate that each occupation-number pattern corresponds to multiple distinguishable microstates. Particularly revealing is the case where each single-particle state contains at most one particle (partition \( (1,1,1,1) \)): we find $6$, $9$, and $10$ distinguishable microstates for parabosons with \( q = 2, 3, 4 \) respectively, compared to exactly 1 for genuine indistinguishable particles.

In conclusion, while parastatistics achieves quantum mechanical consistency through its $U(m)$ invariance (the Hilbert space of parastatistics is not invariant under $S_N$), it fundamentally fails as a statistics for indistinguishable particles. The multiplicity of distinguishable microstates for each occupation-number pattern reveals that para-particles are, in essence, a special class of distinguishable particles, rather than a genuine generalization of quantum statistics.

\subsubsection{A Discussion on the ``Internal'' Degree of Freedom}

One violation of the no-go theorem, i.e.\ instances with $\Omega^K>1$, can be traced to the introduction of an \emph{unmeasurable} ``internal'' degree of freedom. In such cases, the occupation-number pattern no longer uniquely determine the microstates. Allowing such hidden sectors makes various parastatistics constructions (e.g.\ Green-type schemes and related proposals~\cite{sanchez2024reconstruction}) formally realizable, but at the price of imposing strong symmetry constraints on the algebra of admissible observables.

Green's parastatistics provides a canonical example. One enlarges the one-particle Hilbert space to $\mathbb{C}^m \otimes \mathbb{C}^p$,
introducing an auxiliary ``color'' space $\mathbb{C}^p$ with an $SU(p)$ action~\cite{green1953generalized}, and then restricts all physical observables to be $SU(p)$--invariant. This construction generates an internal degeneracy at the Hilbert-space level, yet by design no physical observable can distinguish different color components. It is therefore inconsistent with basic requirements: the hidden $SU(p)$ degeneracy introduces an \emph{unobservable} entropy contribution, so that two gases differing only in $p$ would exhibit distinct thermodynamic responses while being operationally indistinguishable. This Gibbs-type paradox shows that an $SU(p)$-invariant internal-label degeneracy is incompatible with the requirement that physical states be uniquely specified (up to phase) by a complete set of \emph{measurable} quantum numbers.

Conversely, if one \emph{enlarges} the observable algebra so that the ``internal'' degrees of freedom become measurable, then these constructions fall back within the scope of the no-go theorem. The condition $\Omega^K>1$ then directly signals a violation of indistinguishability: Green's scheme no longer defines a new exchange statistics for a single species of identical particles.

\subsection{Parastatistics in Ref.~\cite{wang2025particle}}

In Ref.~\cite{wang2025particle}, a new type of parastatistics was proposed. 
Here, we apply our no-go theorem to explicitly show the violation of the new scheme. 

The commutation relation in general reads~\cite{wang2025particle}
\begin{align}
\psi_{ia}\psi_{jb}^{\dagger} & =\sum_{cd}R_{bd}^{ac}\psi_{jc}^{\dagger}\psi_{id}+\delta_{ab}\delta_{ij},\\
\psi_{ia}^{\dagger}\psi_{jb}^{\dagger} & =\sum_{cd}R_{ab}^{cd}\psi_{jc}^{\dagger}\psi_{id}^{\dagger},\\
\psi_{ia}\psi_{jb} & =\sum_{cd}R_{dc}^{ba}\psi_{jc}\psi_{id}.
\end{align}
with $a,b=1,2,\cdots,m$ denoting the flavors. The $R_{ab}^{cd}$
satisfies the Yang-Baxter equation. Here we focus on the example
by choosing $R_{cd}^{ab}=-\delta_{ac}\delta_{bd}$. 

\subsubsection{Canonical Partition Function: A Fock Space View}

For $R_{cd}^{ab}=-\delta_{ac}\delta_{bd}$, the commutation algebraic relations
read 
\begin{align} \label{eq:wangR}
\psi_{ia}\psi_{jb}^{\dagger} & =-\delta_{ab}\sum_{c}\psi_{jc}^{\dagger}\psi_{ic}+\delta_{ab}\delta_{ij},\\
\psi_{ia}^{\dagger}\psi_{jb}^{\dagger} & =-\psi_{ja}^{\dagger}\psi_{ib}^{\dagger}\label{eq:ab=00003D-ba}\\
\psi_{ia}\psi_{jb} & =-\psi_{ja}\psi_{ib}.
\end{align}
Taking $i=j$ in Eq. (\ref{eq:ab=00003D-ba}), we have the constraint
$\psi_{ia}^{\dagger}\psi_{ib}^{\dagger}=0$, which implies given any
mode $i$, any occupation with two or more particles is forbidden,
even with different flavors. There is no any exclusion between paraparticles
in different modes $i\neq j$. The whole state space is $(m+1)^{N}$-dimensional,
spanned by orthonormal basis states of the form 
\begin{equation}
|\Psi\rangle=\psi_{i_{1}a_{1}}^{\dagger}\psi_{i_{2}a_{2}}^{\dagger}\cdots|0\rangle.
\end{equation}

We consider the single-particle Hamiltonian 
$
H_{1}=-\sum_{\langle ij\rangle}t_{ij}\left(\psi_{ja}^{\dagger}\psi_{ia}+h.c.\right).
$
Now we can calculate the matrix elements of the many-body Hamiltonian
in the Fock space. Towards it, we present two formula, 
\begin{align}
\psi_{kb}|\Psi\rangle & =\psi_{kb}\psi_{i_{1}a_{1}}^{\dagger}\psi_{i_{2}a_{2}}^{\dagger}\cdots|0\rangle\nonumber \\
 & =(-1)^{\#_{k}}\psi_{i_{1}a_{2}}^{\dagger}\psi_{i_{2}a_{3}}^{\dagger}\cdots\psi_{k-1b}^{\dagger}\hat{k}\psi_{k+1a_{k+1}}\cdots|0\rangle,
\end{align}
and 
\begin{align}
\psi_{kb}^{\dagger}|\Psi\rangle & =\psi_{kb}^{\dagger}\psi_{i_{1}a_{1}}^{\dagger}\psi_{i_{2}a_{2}}^{\dagger}\cdots|0\rangle\nonumber \\
 & =(-1)^{\#_{k}}\psi_{i_{1}b}^{\dagger}\psi_{i_{2}a_{1}}^{\dagger}\cdots\psi_{k-1a_{k-2}}^{\dagger}\psi_{ka_{k-1}}^{\dagger}\cdots|0\rangle.
\end{align}
Here $\hat{k}$ means the removal of $\psi_{kb}^{\dagger}$ and
$\#_{k}=\sum_{i<k}n_{i}$ with $n_{i}=\sum_{a}\psi_{ia}^{\dagger}\psi_{ia}$.
By diagonalizing the multi-particle Hamiltonian matrix $H$, we can evaluate the partition
function $Z=\mathrm{Tr}e^{-\beta H}$ based on the many-body spectrum. 

\begin{figure}[t]
\centering
\includegraphics[scale=0.6]{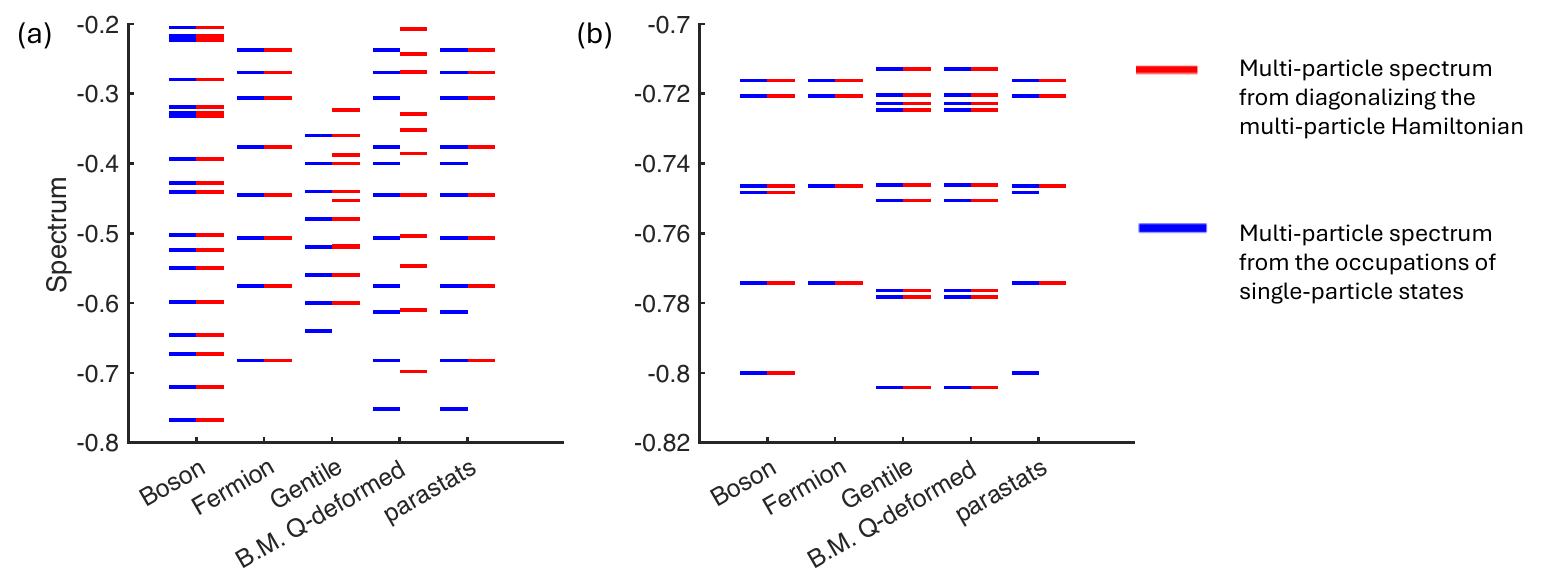}
\caption{Multi-particle spectra for systems obeying Bose, Fermi, Gentile, and parastatistics  for (a) a general Hamiltonian as in Eq.~\eqref{eq:htij} and (b) a diagonal Hamiltonian. Gentile-1 and Gentile-2 refer to the two possible realization schemes.
For a non-diagonal Hamiltonian, only bosonic and fermionic systems remain free.
For a diagonal Hamiltonian, all statistics yield free many-body systems. However, only bosons and fermions lead to non-negative coefficients in the Schur-function expansion of the partition function.
For parastatistics, to construct the many-body spectrum from the single-particle energy states, we simply assume a maximal occupation number to be $m$ which may lead to additional spectral branches.
}
\label{sm:spectrum}
\end{figure}

In Fig.~\ref{sm:spectrum}, we should depict the many-body spectra for a general hopping Hamiltonian 
$
H = \sum_{ij} t_{ij}\psi_{ia}^\dagger \psi_{ja} 
$
and diagonal Hamiltonian 
$
H = \sum_{ia} \epsilon_{ia}\psi^\dagger_{ia} \psi_{ia} 
$. 
As the occupation number for the single-particle states  is complicated, 
we simply assume the maximal occupation number to be $m$ to obtain the many-body spectra from the occupation of the single-particle states, which may account for the extra spectra colored blue.

To check the no-go theorem, we can numerically expand the canonical partition functions for $m=2$ for the diagonal Hamiltonian $H = \sum_i \epsilon_{ia} N_{ia}$ with $N_{ia}= \psi_{ia}^\dagger \psi_{ia}$.
\begin{align}
Z_{m=2}(N=2) & = -\frac{1}{2}s_{(2)} + \frac{3}{2}s_{(1,1)} = \frac{1}{2}m_{(2)} + m_{(1,1)} \label{w2} 
\end{align}
Clearly, the parastatistics in Eq.~\eqref{eq:wangR} fail to neither keep the symmetric group and behave as a proper statistical system. 

\subsubsection{Canonical Partition Function: A Hilbert Space View} \label{wang_model}
We consider a two-particle system in the model with $i=a,b$ and $m=\uparrow,\downarrow$. For a single particle, there are four energy states$|a,\uparrow\rangle$, $|a,\downarrow\rangle$, $|b,\uparrow\rangle$, and $|b,\downarrow\rangle$, with corresponding energy sspectrum, $\{\epsilon_a,\epsilon_a,\epsilon_b,\epsilon_b\}$ while the energy degeneracy is imposed by the commutation relation. In the following, we will show how the parastatistics breaks the no-go theorem.

As a comparison, we illustrate the Hilbert space structure for the boson and fermi systems. The full Hilbert space for two particles is 16-dimensional, spanned by basis vectors of the form $|i,\sigma\rangle \otimes |j,\mu\rangle$, where $i,j \in \{a,b\}$ and $\sigma,\mu \in \{\uparrow,\downarrow\}$.
For conventional bosons and fermions, we need to further choose Hilbert spaces to be symmetric and antisymmetric from this $16$-dimensional space, with dimensions $10$ and $6$ respectively. For example, the basis of antisymmetric subspaces reads:
\begin{equation}
\begin{aligned}
|\psi_1\rangle &= |a,\uparrow\rangle |a,\downarrow\rangle - |a,\downarrow\rangle |a,\uparrow\rangle, \\
|\psi_2\rangle &= |a,\uparrow\rangle |b,\uparrow\rangle - |b,\uparrow\rangle |a,\uparrow\rangle ,\\
|\psi_3\rangle &= |a,\uparrow\rangle |b,\downarrow\rangle - |b,\downarrow\rangle |a,\uparrow\rangle, \\
|\psi_4\rangle &= |a,\downarrow\rangle |b,\uparrow\rangle - |b,\uparrow\rangle |a,\downarrow\rangle, \\
|\psi_5\rangle &= |a,\downarrow\rangle |b,\downarrow\rangle - |b,\downarrow\rangle |a,\downarrow\rangle,\\
|\psi_6\rangle &= |b,\uparrow\rangle |b,\downarrow\rangle - |b,\downarrow\rangle |b,\uparrow\rangle.
\end{aligned}
\end{equation}
It can be verified that the subspace is invariant under the basis transformation of single particle Hilbert space. 

The corresponding Hilbert space $\mathcal{H}$ is $4$-dimensional and exhibits a symmetry that is neither purely symmetric nor antisymmetric. It is spanned by the following basis vectors:
\begin{equation} 
    \begin{aligned}
|\psi_1\rangle &= |a,\uparrow\rangle|b,\downarrow\rangle - |b,\uparrow\rangle|a,\downarrow\rangle, \\
|\psi_2\rangle &= |a,\uparrow\rangle|b,\uparrow\rangle - |b,\uparrow\rangle|a,\uparrow\rangle, \\
|\psi_3\rangle &= |a,\downarrow\rangle|b,\uparrow\rangle - |b,\downarrow\rangle|a,\uparrow\rangle, \\
|\psi_4\rangle &= |a,\downarrow\rangle|b,\downarrow\rangle - |b,\downarrow\rangle|a,\downarrow\rangle.
\end{aligned} \label{eq:basisi2ptwang}
\end{equation}
This subspace supports a representation of the symmetric group $S_2$. The canonical partition function computed within this subspace is $Z=4e^{-\beta(\epsilon_a+\epsilon_b)}=-\frac{1}{2}s_{(2)} + \frac{3}{2}s_{(1,1)}  = \frac{1}{2}m_{(2)}  + m_{(1,1)} $ and coincides exactly with the result obtained via the Fock space method in Eq.~\eqref{w2} for $2$-particle system.

\subsubsection{Inconsistent with both Quantum Mechanics and Statistics}\label{fracture_discussion}
Crucially, the $s$-function expansion in Eq.~\eqref{w2} contains negative coefficients, while the $m$-function coefficients are neither strictly $0$ nor $1$. This demonstrates unambiguously that such statistics are fundamentally incompatible with both the basis independence and the principle of indistinguishability in statistical mechanics. 

\paragraph{The Hilbert space is not close under $U(m)$ transformation on single-particle Hilbert space} We show that the Hilbert space with basis defined in Eq.~\eqref{eq:basisi2ptwang} \textit{does not} support a representation of the full unitary group $U(4)$, violating basis independence. 

The single-particle Hilbert space can be spanned by $\{|a,\uparrow\rangle,|a,\downarrow\rangle,|b,\uparrow\rangle,|b,\downarrow\rangle\}$. We can consider a new basis for the single-particle space, $\{|a,\uparrow\rangle,\frac{1}{\sqrt{2}}(|a,\downarrow\rangle + |b,\uparrow\rangle), \frac{1}{\sqrt{2}}(|a,\downarrow\rangle - |b,\uparrow\rangle) ,|b,\downarrow\rangle\}$, which is obtained by a unitary transformation $T$ in $U(4)$ on the old basis. We can now check how $T'$ acts on a basis in two-particles Hilbert space.
When $T$ acts on the basis vector $|\psi_2\rangle$ in Eq.~\eqref{eq:basisi2ptwang}, we obtain:
\begin{align}
T'|\psi_2\rangle &\equiv T|a,\uparrow\rangle \otimes T|b,\uparrow\rangle - T|b,\uparrow\rangle \otimes T|a,\uparrow\rangle \notag \\
&= \frac{1}{\sqrt{2}}(|a,\uparrow\rangle|a,\downarrow\rangle - |a,\downarrow\rangle|a,\uparrow\rangle) + \cdots.
\end{align}
Clearly, the transformed state $T'|\psi_2\rangle$ does not belong to the original Hilbert space spanned by $\{\psi_1,\psi_2,\psi_3,\psi_4\}$ in Eq.~\eqref{eq:basisi2ptwang}. This demonstrates that the Hilbert space $\mathcal{H}$ is not invariant under the unitary group $U(m)$. Therefore, the physical description of this statistics depends on the choice of basis in the single-particle Hilbert space. 

The space $\operatorname{span}\{|\psi_1\rangle, |\psi_2\rangle, |\psi_3\rangle, |\psi_4\rangle\}$ forms an $S_N$-invariant subspace which is a reducible representation of $S_2$. This representation decomposes into the direct sum of the identity representation (appearing once) and the fully antisymmetric representation (appearing three times).

\paragraph{Counting microstates: interpreting fractional microstate numbers}
From the $m$-function expansion of the canonical partition functions in Eq.~\eqref{w2}, we see that in the two-particle case the configuration with both particles in the same level appears with coefficient $1/2$, and in the three-particle case the configuration with all three particles in the same level also carries a coefficient $1/2$. In the Haldane–Wu framework \cite{haldane1991fractional,wu1994statistical} (see Sec.~\ref{hws}; see also the microscopic analysis by Chaturvedi et al.~\cite{chaturvedi1997microscopic}), such fractional coefficients are interpreted not as “half a microstate” but as \emph{statistical weights} attached to each occupation-number pattern $(\lambda)$. From the viewpoint of conventional quantum mechanics, however, this reinterpretation allowing non-integer amounts to abandoning the strict microstate-counting interpretation of the partition function. On the other hand, the Haldane--Wu framework~\cite{haldane1991fractional,wu1994statistical}
is widely regarded as an effective statistical description, in which fractional occupation-number-pattern weights do not correspond to literal microstate multiplicities, but rather provide an
effective encoding of interaction effects or statistical correlations
\cite{murthy1994thermodynamics,isakov1994statistical}.

\subsubsection{Why Parastatistics in Ref.~\cite{wang2025particle} Fails}
A multiplicity of microstates associated with a given occupation-number pattern must either violate particle indistinguishability, or require the introduction of additional information such that occupation-number pattern alone no longer fully specify the microstate. Green’s parastatistics introduces completely unobservable internal degrees of freedom, yet counts them as distinct microstates; this inevitably leads to a Gibbs-like paradox. One possible resolution is to introduce a mechanism that renders the additional degrees of freedom operationally accessible. However, unlike anyonic systems—where the extra information is encoded in topological degeneracy—parastatistics in Ref.~\cite{wang2025particle} still relies on single-particle internal degrees of freedom to construct a globally measurable structure. We prove that such a construction unavoidably destroys the consistency of the many-body Hilbert space. Moreover, it reduces occupation-number pattern counting to mere statistical weights, thereby forfeiting its interpretation as genuine microstate counting.

\subsection{Haldane-Wu's Model: Semion Statistics}\label{hws}
Here we consider the semion statistics as a representative example in Haldane-Wu's model with $g=1/2$.
The canonical partition function of such system for $5$ particle system is given by \cite{chaturvedi1997microscopic}
\begin{align}\label{semm}
Z &= \frac{1}{3}m_{(2,2,1)}(x_1,x_2,\ldots) + \frac{1}{2}m_{(2,1,1,1)}(x_1,x_2,\ldots)+m_{(1,1,1,1,1)}(x_1,x_2,\ldots).
\end{align}
We can express it in terms of $s$-function, which reads
\begin{align}\label{sems}
Z &= \frac{1}{3}s_{(2,2,1)}(x_1,x_2,\ldots)-\frac{1}{6}s_{(2,1,1,1)}(x_1,x_2,\ldots).
\end{align}

The Semion statistics also reveals fundamental limitations at both the statistical level (microstate indistinguishability) and the quantum-mechanical level (basis independence), as evidenced by the non-standard form of the partition function with fractional and negative coefficients in the $m$-function and $s$-function expansions given in Eqs.~\eqref{semm} and \eqref{sems}.

\subsection{Quon Algebra and Its Deformation} \label{section:quon}

\subsubsection{Quon Algebra}
The quon algebra \cite{greenberg1991particles,greenberg1993quons} represents a systematic and continuous deformation of particle statistics, interpolating between Boson and Fermi statistics by introducing a real parameter $q$. Its defining relation is 
\begin{equation}\label{quon-normal}
    a_{i}^{}a_{j}^{\dagger}-qa_{j}^{\dagger}a_{i}^{}=\delta_{ij},
\end{equation}
where $i,j$ label discrete modes and $q\in\mathbb{R}$. Here $a^\dagger$ is the adjoint of $a$.  they are not Hermitian conjugated pair when $q\neq \pm1$. 

Quon systems display several characteristics \cite{2008arXiv0805.0285G}: \textbf{1)}. The Fock space constructed from quon operators is positive definite when $-1\leq q\leq1$. \textbf{2)}, For value $q^{2}\neq1$, there are no exchange relations among the annihilation operators $a_{i}$, $a_{j}$ or among the creation operators $a_{i}^{\dagger}$, $a_{j}^{\dagger}$, Thus, for a fixed sequence of indices $1,2,\cdots,n$, all $n!$ states of the form $a_{i_{1}}\cdots a_{i_{n}}|0\rangle$ are linearly independent if no quantum numbers repeat. \textbf{3)}. The number operator in the quon theory must be expressed as an infinite series in creation and annihilation operators, with coefficients that grow increasingly complex and diverge as $q^{2}\rightarrow1$, reflecting the loss of a simple quantized spectrum in the Bose/Fermi limits. \textbf{4)} Quon models are intrinsically nonlocal.

\subsubsection{Quon and Distinguashable Particles}

Here, we focus on the physically relevant range $-1 < q < 1$, noting that the special case $q = 0$ corresponds to what is known as infinite statistics~\cite{greenberg1990example}.

The creation and annihilation operators defined in Eq.~\eqref{quon-normal} \emph{do not} form either symmetric or antisymmetric subspaces, and the $N$-particle Hilbert space remains the full tensor product $\mathcal{V}_N = V^{\otimes N}$. The deformation parameter $q$ enters the theory by altering the definition of the inner product in the Fock space. Nonetheless, the trace of an operator, and hence the partition function, remains independent of $q$. This can be seen from a direct calculation: $\langle 0 | a_i a_j a_j^{\dagger} a_i^{\dagger}|0\rangle = 1 + q \delta_{ij}$ with $|0\rangle$ the vacuum. To construct an orthonormal basis, the two-particle states must be normalized as $\frac{1}{\sqrt{1 + q \delta_{i,j}}} a_j^{\dagger} a_i^{\dagger}|0\rangle$. Thus, the canonical partition function for an \emph{ideal} quon gas coincides exactly with that of completely distinguishable particles, which is independent of the deformation parameter $q$, as established in Ref.~\cite{goodison1994canonical}. 

In the language of symmetric functions, the canonical partition function for infinite statistics, or the Maxwell-Boltzmann statistics, can be written as
\begin{align} \label{Z_quon}
Z_{\text{Infinity}}(\beta,N) &= \sum_{I}^{P(N)} \dim(V_{S_N}^I) s_I(x_1,x_2,\ldots) \\
&= \sum_{I}^{P(N)} \frac{N!}{\prod_{j=1}^N \lambda_{I,j}!} m_I(x_1,x_2,\ldots) = \left(\sum_i x_i\right)^N,
\end{align}
which coincides with the partition function for completely distinguishable particles.
For instance, for $N=3$, we have
\begin{align}
Z_{\text{Infinity}}(\beta,N=3) &= m_{(3)}(x_1,x_2,\ldots) + 3m_{(2,1)}(x_1,x_2,\ldots) + 6m_{(1,1,1)}(x_1,x_2,\ldots) \\
  &= s_{(3)}(x_1,x_2,\ldots) + 2s_{(2,1)}(x_1,x_2,\ldots) + s_{(1,1,1)}(x_1,x_2,\ldots).
\end{align}
This is equivalent to the Maxwell-Boltzmann statistics while the coefficients in both the $m$-function, $\Omega^I= \frac{N!}{\prod_{j=1}^N \lambda_{I,j}!} $, and $s$-function expansions indicate all particles are distinguishable. The core conclusion is that despite the $q$-dependent mechanical structure of the Quon system in Eq.~\eqref{quon-normal}, it is statistically indistinguishable from a system of distinguishable particles. This reveals its fundamental inadequacy as a description of indistinguishable quantum particles.

\subsubsection{Distinguishable Particle with Maximum Occupation Number}
While the deformation in Eq.~\eqref{quon-normal} alters the inner product of the Hilbert space for a free quon gas, we now show that generalized commutation relations can go further—imposing an upper bound on the occupation number of single-particle states. 
We now examine the case of quons subject to a maximum occupancy constraint, 
\begin{equation}\label{onlymax}
    a_i^{} a_j^\dagger - \omega a_j^\dagger a_i = \omega ^{-\mathcal N_i}\delta_{ij},
\end{equation}
where $\omega = e^{i\pi/(q+1)}$.  By assuming a unique vacuum $|0\rangle$ such at $a_i|0\rangle =0$ $\forall i$. The particle number operator $N_i$ satisfies $N_i |n_i\rangle = n_i |n_i\rangle$, where $|n_i\rangle$ represents a state with $n_i$ particles in mode $i$. The creation and annihilation operators act on these states according to:
\begin{equation}
a_i^\dagger |n_i\rangle = \sqrt{f(n_i+1)} |n_i+1\rangle, \quad a_i |n_i\rangle = \sqrt{f(n_i)} |n_i-1\rangle, 
\end{equation}
where $f(n_i)$ is a real-valued function satisfying $a_i^\dagger a_i |n_i\rangle = f(n_i) |n_i\rangle$.

To determine the functional form of $f(n)$, we apply the algebraic relation $(a_i a_i^\dagger - \omega a_i^\dagger a_i)|n_i\rangle = \omega^{-N_i}|n_i\rangle$ with $\omega = e^{i\pi/(q+1)}$. Substituting the operator actions yields:
\begin{equation}
f(n_i+1) - \omega f(n_i) = \omega^{-n_i}.
\end{equation}
This recurrence relation, combined with the initial condition $f(0) = 0$ (from $a_i|0\rangle = 0$), has the unique solution:
\begin{equation}
f(n) = \frac{\sin\left(\frac{n\pi}{q+1}\right)}{\sin\left(\frac{\pi}{q+1}\right)}.
\end{equation}
One can readily verify that $f(q+1) = 0$, which implies that the state $|q+1\rangle$ has zero norm and is therefore unphysical. Consequently, the maximum occupation number for any single-particle state is $q$, establishing an exclusion principle analogous to Gentile statistics.

Consider a three-particle system where each particle occupies a four-dimensional single-particle Hilbert space ($m=4$), with the deformation parameter $q=2$ enforcing a maximum occupation number. In the absence of any commutation relations between $a_i$ and $a_j$, the Fock space has the structure of the full tensor product space $V^{\otimes 3}$, with $64$ basis states. Imposing the maximum occupation number $q=2$ removes the four states of the form $(a_i^\dagger)^3 |0\rangle$, where $i=1,\ldots,4$. The resulting allowed subspace is $60$-dimensional.

The canonical partition function can thus be computed directly as:
\begin{align}
Z^{m=4,q=2}(\beta,N=3) &= 0\cdot m_{(3)}(x_1,x_2,\ldots) + 3\cdot m_{(2,1)}(x_1,x_2,\ldots) + 6\cdot m_{(1,1,1)}(x_1,x_2,\ldots) \\
  &= 3s_{(2,1)}(x_1,x_2,\ldots) .
\end{align}
This provides the canonical partition function of a system of distinguishable particles subject to a maximal-occupancy constraint. Although such a construction violates the indistinguishability principle. Moreover, the Hilbert space is $U(m)$ invariant (since the coefficient of $s$-function are all non-negative) but not $S_N$ invariant (the coefficient is not $2$, the dimension of irrep of $S_N$ labeled by $(2,1)$).

\subsubsection{$Q$-Deformed Algebras}
Alongside the quon algebra, further developments have explored a variety of generalizations, including several forms of q-deformed algebras. As one example, a q-deformed algebra of creation and annihilation operators has been proposed in which the operators carry ordered indices and the deformation parameter is allowed to be complex.:
\begin{align}
& a_{i}^{}a_{i}^{\dagger}-q a_{i}^{\dagger}a_{i}^{}	= \Lambda(N_i),  \label{inner} \\
 & a_{i}a_{j}-s a_{j}a_{i}	= a_{i}^{}a_{j}^\dagger -s a_{j}^\dagger a_{i}^{}=0, \quad i<j \label{restrict}
\end{align}
with $q,s\in\mathbb{C}$ and $\Lambda(N)$ being the function of the particle number operator. In particular, $a^\dagger$ is the Hermitian conjugate of $a$. For a generic $q$, there is no limitation on the maximal occupation which resembles the Bose-Einstein distribution while specific forms of $q$ and $\Lambda(N_i)$ can lead to a finite maximal occupation number \cite{dai2012calculating}. In the following, we will focus on characteristic examples.

\paragraph{Gentile Statistics from Deformed Quon algebra} \label{sec_quon}

We can examine Gentile statistics from a quantum-mechanical perspective using operator realizations. Specifically, we consider whether one can construct creation and annihilation operators that enforce the Gentile maximum-occupation constraint while preserving the required statistical consistency, namely, a one-to-one correspondence between occupation-number configurations and microscopic quantum states. Several algebraic constructions of Gentile creation and annihilation operators have been proposed in the literature \cite{dai2012calculating}. In what follows, we analyze one example of these constructions and assess their consistency with the basic requirements of a quantum statistics.

One of the possible second quantization realization of the maximal occupation number $q\in \mathbb Z$ is 
\begin{equation}
    a^{\dagger}a  =\frac{\sin\frac{N\pi}{q+1}}{\sin\frac{\pi}{q+1}}, \quad
aa^{\dagger}  =\frac{\sin\frac{(N+1)\pi}{q+1}}{\sin\frac{\pi}{q+1}},
\end{equation}
where $q$ is the maximal occupation of an energy state and $N$
is the particle number operator. We remark that, unlike in the bosonic/fermionic oscillator, the number operator $N$ enters nonlinearly here; the ladder structure is correspondingly more intricate. The quantization
condition reads 
\begin{equation}
aa^{\dagger}-\cos\frac{\pi}{q+1}a^{\dagger}a=\cos\frac{N\pi}{q+1}.
\end{equation}

For a multimode system with a $m$-dimensional single-particle Hilbert space, we can consider the following commutation relation,
\begin{align} \label{eq:Gentile_commutative}
    & a_i^{}a^{\dagger}_i-\cos\frac{\pi}{q+1}a_i^{\dagger}a_i^{}=\cos\frac{N\pi}{q+1} ,\\
&  a_{i}a_{j}  - sa_{j}a_{i} =  a_{i}^{}a_{j}^\dagger  - sa_{j}^\dagger a_{i}^{} = 0,\quad i<j ,
\end{align}
where for $s=1$ a bosonic commutation relation is set between different modes. In practice, the modes $i,j=1,\cdots,m$ can be interpreted as therder of the lattice sites. 

To construct the Fock space, we start with the vacuum $|0\rangle$ by assuming $a|0\rangle=0$ for the single mode.
Then we can construct the $n$-particle state $|n\rangle$ ($n=0,1,\cdots,q$):
\begin{align}
|n\rangle & =\frac{1}{\sqrt{\mathcal{N}_{n}}}\left(a^{\dagger}\right)^{n}|0\rangle,
\end{align}
with $\mathcal{N}_{n}$ being the normalization factor $\mathcal{N}_{n}=\langle0|a^{n}\left(a^{\dagger}\right)^{n}|0\rangle$.
The state $|n\rangle$ contains $n$ particles: $N|n\rangle=n|n\rangle$ with $N$ the particle number operator.
By noting 
\begin{align}
a\left(a^{\dagger}\right)^{n} 
 =\cos\frac{N\pi}{q+1}\left(a^{\dagger}\right)^{n-1}+\cos\frac{\pi}{q+1}a^{\dagger}a\left(a^{\dagger}\right)^{n-1},
\end{align}
one can derive the iterative relation 
\begin{align}
\mathcal{N}_{n} & =[n]\mathcal{N}_{n-1},
\end{align}
with the notation,
\begin{equation}
[n]=\sum_{j=0}^{n-1}\cos^{j}\frac{\pi}{q + 1}\cos\frac{(n-1-j)\pi}{q + 1}.
\end{equation}
Thus, we obtain the normalization factor 
$\mathcal{N}_{n}=[n]!\equiv\prod_{j=1}^{n}[n]$.
A factorial notation is borrowed here as $\mathcal{N}_{n}$ recovers the
Boson/Fermion case when $q=1$ or $q=\infty$. Simultaneously,
we present several useful relations 
\begin{align}
\langle n-1|a|n\rangle & =\sqrt{[n]},\\
\langle n+1|a^{\dagger}|n\rangle & =\sqrt{[n+1]},\\
\langle n|a^{\dagger}a|n\rangle & =[n].
\label{eq:<ada>}
\end{align}
One can find that $a^{\dagger}a$ may not behave as the
particle number operator as shown in Eq.~(\ref{eq:<ada>}) as $[n]$ can be different from $n$.

For the basis in the multi-particle Fock space, the ordering of creation
operators defines distinct basis conventions. In this case, we choose the order 
\begin{equation}
|n_{1}n_{2}\cdots\rangle\equiv\frac{\left(a_{1}^{\dagger}\right)^{n_{1}}}{\sqrt{[n_{1}]!}}\frac{\left(a_{2}^{\dagger}\right)^{n_{2}}}{\sqrt{[n_{2}]!}}\frac{\left(a_{3}^{\dagger}\right)^{n_{3}}}{\sqrt{[n_{3}]!}}\cdots|0\rangle,
\end{equation}
which satisfy the orthonormal conditions 
$\langle n_{1}^{\prime}n_{2}^{\prime}\cdots|n_{1}n_{2}\cdots\rangle=\delta_{n_{1}n_{1}^{\prime}}\delta_{n_{2}n_{2}^{\prime}}\cdots$.

We consider a Hamiltonian of quadratic form without interaction,
\begin{equation} \label{eq:htij}
H_1 =-\sum_{\langle ij\rangle}t_{ij}\left(a_{i}^{\dagger}a_{j}+a_{j}^{\dagger}a_{i}\right),
\end{equation}
with the hopping integral $t_{ij}$ between $i$ and $j$. Given the basis $|\cdots n_{i}\cdots n_{j}\cdots\rangle$, we have the non-zero hopping matrix elements:
\begin{itemize}
\item For $i<j$, we have ($s=1$)
\begin{align}
\langle\cdots n_{i}+1\cdots n_{j}-1\cdots|a_{i}^{\dagger}a_{j}|\cdots n_{i}\cdots n_{j}\cdots\rangle & =s^{-\sum_{k<i}n_{k}}s^{\sum_{k<j}n_{k}}\sqrt{[n_{i}+1][n_{j}]},
\end{align}
\item For $i>j$, we have 
\begin{equation}
\langle\cdots n_{j}+1\cdots n_{i}-1\cdots|a_{i}^{\dagger}a_{j}|\cdots n_{j}\cdots n_{i}\cdots\rangle=s^{1-\sum_{k<i}n_{k}}s^{\sum_{k<j}n_{k}}\sqrt{[n_{i}+1][n_{j}]}.
\end{equation}
\end{itemize}
Here we keep $s$ as in a general case $s$ can induce a non-trivial phase string. 
Then we can diagonalize the multi-particle Hamiltonian to get the partition function $Z=\mathrm{Tr}e^{-\beta H}$ and then expand it in terms of $m$-function and $s$-functions respectively.

\paragraph{Gentile Statistics from Biedenharn–Macfarlane Type $q$-Oscillators} \label{BMquon}
We consider the Biedenharn–Macfarlane type $q$-oscillators, which provide a concrete realization of this algebraic framework among Refs.~\cite{1995hep.th...12083B,1999PrPNP..43..537B,1989JPhA...22L.873B,1989JPhA...22L.983S, 1976JMP....17..524A,1991JPhA...24L.711C,1991JPhA...24L.775J,Ohnuki1982,1991JPhA...24L.591O,1990JPhA...23L1019F,1991PhRvA..44.8020C,1990CMaPh.127..129H,1991JPhA...24L.179F,Gangopadhyay1991,Frappat1991,Parthasarathy1991,Viswanathan1992}. We consider the following algebraic relations,
\begin{align}
 & a_{i}a_{i}^{\dagger}-e^{i\frac{\pi}{q+1}}a_{i}^{\dagger}a_{i}=e^{-i\frac{N_{i}\pi}{q+1}}, \\
&  a_{i}a_{j}  - s a_{j}a_{i} = a_{i}a_{j}^\dagger  - s a_{j}^\dagger a_{i}= 0 \quad i<j.
\end{align}
where for simplicity we can choose $s= \pm1$ while others $s$ may describe nontrivial braiding which will not change our conclusions.
We highlight that the particle number $N_i$ generally does not have the simple expression as the case of boson and fermion.

With the vacuum $|0\rangle$ with $a|0\rangle=0$, we can construct
the $n$-particle state $|n\rangle$($n=0,1,\cdots,q$): 
$
|n\rangle=\frac{\left(a^{\dagger}\right)^{n}}{\sqrt{[n]!}}|0\rangle,
$
with 
$
[n]!  \equiv\prod_{k=1}^{n}[k]=\prod_{k=1}^{n}\frac{\sin\left(\frac{k\pi}{q+1}\right)}{\sin\left(\frac{\pi}{q+1}\right)}$.
It is easy to check $[n]!=n!$ when $q=1$, or $\infty$. 
Similarly, we have the multi-particle basis for the Fock space, 
\begin{equation}
|n_{1}n_{2}\cdots\rangle\equiv\frac{\left(a_{1}^{\dagger}\right)^{n_{1}}}{\sqrt{[n_{1}]!}}\frac{\left(a_{2}^{\dagger}\right)^{n_{2}}}{\sqrt{[n_{2}]!}}\frac{\left(a_{3}^{\dagger}\right)^{n_{3}}}{\sqrt{[n_{3}]!}}\cdots|0\rangle,
\end{equation}
with the orthonormal conditions 
$
\langle n_{1}^{\prime}n_{2}^{\prime}\cdots|n_{1}n_{2}\cdots\rangle=\delta_{n_{1}n_{1}^{\prime}}\delta_{n_{2}n_{2}^{\prime}}\cdots$.

Given the basis $|\cdots n_{i}\cdots n_{j}\cdots\rangle$, we have
the elements by denoting $s=\pm 1$:
\begin{itemize}
\item For $i<j$, we have 
\begin{align}
\langle\cdots n_{i}+1\cdots n_{j}-1\cdots|a_{i}^{\dagger}a_{j}|\cdots n_{i}\cdots n_{j}\cdots\rangle & =s^{\sum_{k<i}n_{k}}s^{\sum_{k<j}n_{k}}\sqrt{[n_{i}+1][n_{j}]},
\end{align}
\item For $i>j$, we have 
\begin{equation}
\langle\cdots n_{j}+1\cdots n_{i}-1\cdots|a_{i}^{\dagger}a_{j}|\cdots n_{j}\cdots n_{i}\cdots\rangle=s^{1-\sum_{k<i}n_{k}}s^{\sum_{k<j}n_{k}}\sqrt{[n_{i}+1][n_{j}]}.
\end{equation}
\end{itemize}
Then, we can construct the multi-particle Hamiltonian matrix.

The above two examples of q-deformed algebra in Sec.~\ref{sec_quon} and Sec.~\ref{BMquon} give rise to a limitation on the maximum occupation number $q$.

\begin{table}[t]
\centering
\caption{Various statistical schemes with principles: (i) single-particle-basis independence, and (ii) indistinguishable-particle microstate counting (unique $\Omega$ per occupation-number pattern). Only bosons and fermions satisfy all two. Here, $\times$/$\checkmark$ means that the case is parameter-dependent.}
\label{tab:qsit-comparison}
\vspace{6pt}
\begin{tabular}{lcccc}
\toprule
\multicolumn{2}{c}{} & \multicolumn{2}{c}{\shortstack{\textbf{Indistinguashability}}}  &\multicolumn{1}{c}{\textbf{Criteria}} \\
\cmidrule(lr){3-4}
\textbf{Scheme} & \textbf{\shortstack{Hilbert\\ space}} & \shortstack{\textbf{Basis Indep.}}& \textbf{Unique $\Omega$} &   \\ 
\midrule
Boson  &$\mathrm{Sym}^N(V)$     & $\checkmark$ & $\checkmark$ & $\checkmark$ \\
\cmidrule(lr){2-2}
Fermion&    $\wedge^N V$     & $\checkmark$ & $\checkmark$ & $\checkmark$ \\
\cmidrule(lr){2-2}
Gentile \cite{gentile1940itosservazioni} &\shortstack{$\mathrm{Sym}^N(V)$\\with maximum \\ occupation \\number}   & $\times$ & $\checkmark$ & $C<0$  \\
\cmidrule(lr){2-2}
\shortstack{Greenberg's \\Quon \cite{greenberg1993quons,greenberg1991particles} }& $V^{\otimes N}$ & $\checkmark$ & $\times$ & $\Omega >1$ \\
\cmidrule(lr){2-2}
\shortstack{Green's\\ parastatistics \cite{green1953generalized}}&$\bigoplus\left( V^I_{U(m)}\right) $   & $\checkmark$ & $\times$ & $\Omega >1$ \\
\cmidrule(lr){2-2}
Immanons \cite{tichy2017extending} &       $ V^I_{U(m)}$                  &  $\checkmark$  & $\times$ & $\Omega >1$  \\
\cmidrule(lr){2-2}
\shortstack{Biedenharn--\\Macfarlane \cite{biedenharn1989quantum} } &  \shortstack{With maximum \\ occupation \\number}  & $\times$ & $\times$/$\checkmark$  &  \shortstack{$C<0$ or \\ additionally $\Omega >1$} \\
\cmidrule(lr){2-2}
Parastatistics \cite{wang2025particle}& - & $\times$ & $\times$ & Fractional $\Omega$  and $C<0$ \\
\cmidrule(lr){2-2}
 
 \shortstack{Haldane-Wu's Model:\\
 Semion Statistics \cite{chaturvedi1997microscopic} } &    -         & $\times$ & $\times$ & Fractional $\Omega$ and $C<0$ \\
\cmidrule(lr){2-2}

Jack polynomials \cite{chaturvedi1997interpolations}    &      -        &$\checkmark$/ $\times$ & $\times$ & \shortstack{ $\Omega >1$ or \\ additionally $C<0$ }\\

\bottomrule
\end{tabular}
\end{table}

\paragraph{Single-particle basis dependence} Numerically, by constructing the many-body Hamiltonian given a general quadratic Hamiltonian, we find the system is no longer free as the many-body spectra do not match the spectra from the occupations of  single-particle states. The inconsistency can be understood from the generalized Pauli exclusion principle which leads to the breakdown of the single-particle-basis independence in the multi-particle Fock space. Alternatively, the Fock space does not keep invariant under the transformation of $U(m)$ group. By contrast, a diagonal Hamiltonian of the form $H_1= \sum_i\epsilon_i N_i$ corresponds to a free system where $N_i$ is the particle number operator $N_i|n_i\rangle = n_i |n_i\rangle$. However, a negative $C^I$ obtained from numerical calculations indicates the violation of the no-go theorem (violating the basis independence).

\subsection{Other Generalizations of Quantum Statistics}
To illustrate the broader landscape of proposed statistical generalizations, we examine several representative examples through the lens of our no-go theorem. For instance, Jack polynomials \cite{chaturvedi1997interpolations} exemplify approaches that employ symmetric polynomials to generalize quantum statistics \cite{zhou2022unified}. Immanons constitute another paradigm that utilizes higher-dimensional representations of permutation groups to implement intermediate statistics \cite{tichy2017extending}. Furthermore, generalized statistics frameworks that permit distinct occupation numbers for different states \cite{dai2004gentile} are explicitly excluded by the analysis in Section~\ref{sec_example}, due to the arbitrary selection of coefficients $\Omega$.

\paragraph{Jack Polynomials}
The approach using Jack polynomials \cite{chaturvedi1997interpolations} provides an interpolation between particle through a continuous parameter $\alpha$. For the partition $(2,1)$, the partition function takes the form:
\begin{align}
Z_{Jack}(\beta,N=3) &= J_{(2,1)}^{(\alpha)} = (2+\alpha)m_{(2,1)} + 6m_{(1,1,1)} \\
  &= (2+\alpha)s_{(2,1)} + 2(1-\alpha)s_{(1,1,1)}
\end{align}
The $\alpha$-dependent coefficients demonstrate how this construction attempts to bridge between statistical classes.

\paragraph{Immanons} Immanons \cite{tichy2017extending} provide yet another alternative, characterized by their specific combinatorial structure. For partition $(2,1)$, we obtain:
\begin{align}
Z_{Immanons}(\beta,N=3) &= m_{(2,1)} + 2m_{(1,1,1)} \\
  &= s_{(2,1)}
\end{align}
This simplified pattern reflects the particular representation-theoretic construction underlying Immanons.

These explicit expressions reveal the fundamental shortcomings common to such constructions: they invariably fail to satisfy either the consistency requirements of statistical counting (as manifested in the $m$-function coefficients) or the quantum mechanical constraints on Hilbert space structure (as reflected in the $s$-function expansion). In each case, the mathematical structure proves incompatible with the fundamental principles established in our no-go theorem.

\paragraph{More on Q-deformed Algebra} \label{q-deform-boson}
In the literature, various q-deformed algebras have been proposed as candidates for intermediate particle exchange. It is well known that naive q-deformed commutation relations are not preserved under general  basis transformations of the one-particle Hilbert space. This basis dependence has been widely recognized as a structural issue. For example, a wide class of generalized q-deformed algebras (see e.g., Refs.~\cite{1995hep.th...12083B,1999PrPNP..43..537B,1989JPhA...22L.873B,1989JPhA...22L.983S,1976JMP....17..524A,1991JPhA...24L.711C,1991JPhA...24L.775J,Ohnuki1982,1991JPhA...24L.591O,1990JPhA...23L1019F,1991PhRvA..44.8020C,1990CMaPh.127..129H,1991JPhA...24L.179F,Gangopadhyay1991,Frappat1991,Parthasarathy1991,Viswanathan1992}) leads to Gentile statistics~\cite{dai2012calculating}. These schemes are directly excluded by our no-go theorem, as the violation of $C^I$ in the canonical partition function implies a breakdown of basis independence. Thus, the q-deformed algebras motivate the embedding of q-oscillator algebras into quantum-group frameworks based on $U_q(\mathfrak{g})$, Hecke algebras, or $R$-matrix structures~\cite{leblanc1993r,macfarlane1989q,kulish1993quantum}.

\end{document}